\documentclass{lmcs} 
\pdfoutput=1
\usepackage[utf8]{inputenc}

\usepackage{lastpage}
\lmcsdoi{21}{2}{3}
\lmcsheading{}{\pageref{LastPage}}{}{}%
{Mar.~01,~2024}{Apr.~08,~2025}{}

\usepackage{algorithm}
\usepackage{algpseudocodex}

\usepackage{graphicx}
\usepackage{wrapfig}
\usepackage{tikz}
\newcommand{\set}[1]{\{ #1 \}}
\newcommand{\PSPACE}{\mathrm{PSPACE}}
\newcommand{\PTIME}{\mathrm{PTIME}}

\newcommand{\EXP}{\mathrm{EXPTIME}\xspace}

\newcommand{\M}{\mathcal{M}}

\newcommand{\play}{\pi}

\newcommand{\G}{\mathcal{G}}

\newcommand{\zug}[1]{\langle #1 \rangle}

\newcommand{\stam}[1]{}
\usepackage{xspace}

\newcommand{\PO}{Player~$1$\xspace}
\newcommand{\PT}{Player~$2$\xspace}
\newcommand{\PLi}{Player~$i$\xspace}
\newcommand{\PLni}{Player~$-i$\xspace}
\renewcommand{\ni}{-i\xspace}
\renewcommand{\S}{\mathcal{S}}%{{\cal S}}
\renewcommand{\P}{\mathcal{P}}%{{\cal P}}
\newcommand{\LAK}{\text{Lock \& Key}\xspace}
\newcommand{\key}{k}%\textsc{Key}\xspace}
\newcommand{\lock}{\ell}%{\textsc{Lock}\xspace}
\newcommand{\Nat}{\mathbb{N}\xspace}
\newcommand{\A}{\mathcal{A}\xspace}
\newcommand{\pos}{i}

\usetikzlibrary{automata,arrows}
\begin{document}

\title[A game of Pawns]{A game of Pawns}

%
% and a list of author information of the form
%
\author[G. Avni]{Guy Avni\lmcsorcid{0000-0001-5588-8287}}[a]
\address{Department of Computer Science, University of Haifa, Haifa, Israel}
\email{gavni@cs.haifa.ac.il}
\thanks{Partially supported by ISF grant no. 1679/21.}

\author[P. Ghorpade]{Pranav Ghorpade\lmcsorcid{0009-0001-0421-4490}}[b]
\address{The University of Sydney, Sydney, Australia}
\email{pranav.ghorpade@sydney.edu.au}

% \thanks{thanks 2}

\author[S. Guha]{Shibashis Guha\lmcsorcid{0000-0002-9814-6651}}[c]
\address{Tata Institute of Fundamental Research, Mumbai, India}
\email{shibashis@tifr.res.in}
\thanks{Partially supported by the SERB grant no. SRG/2021/000466.}

%
% The \author, \address and \email fields are mandatory.
%
% The \thanks fields are optional.  They appear
% in footnotes on the title page, the addresses and email information
% is relegated to the top of the article.  The optional arguments to
% the \title and \author macros determine a running head on the odd
% and even pages, respectively.

\begin{abstract}
We introduce and study {\em pawn games}, a class of two-player zero-sum turn-based graph games. A turn-based graph game proceeds by placing a token on an initial vertex, and whoever {\em controls} the vertex on which the token is located, chooses its next location. This leads to a path in the graph, which determines the winner. Traditionally, the control of vertices is predetermined and fixed. 
The novelty of pawn games is that control of vertices changes dynamically throughout the game as follows. 
Each vertex of a pawn game is {\em owned} by a {\em pawn}. In each turn, the pawns are partitioned between the two players, and the player who {\em controls} the pawn that owns the vertex on which the token is located, chooses the next location of the token. 
Control of pawns changes dynamically throughout the game according to a fixed mechanism.
Specifically, we define several {\em grabbing}-based mechanisms in which control of at most one pawn transfers at the end of each turn. We study the complexity of solving pawn games, where we focus on reachability objectives and parameterize the problem by the mechanism that is being used and by restrictions on pawn ownership of vertices. 
On the positive side, even though pawn games are exponentially-succinct turn-based games, we identify several natural classes that can be solved in PTIME. 
On the negative side, we identify several $\EXP$-complete classes, where our hardness proofs are based on a new class of games called \LAK games, which may be of independent interest.
\end{abstract}

\maketitle

\section{Introduction}
Two-player zero-sum {\em graph games} constitute a fundamental class of games~\cite{AG11} with applications, e.g., in {\em reactive synthesis}~\cite{PR89}, multi-agent systems~\cite{AHK02}, a deep connections to foundations of logic \cite{Rab69}, and more.
A graph game is played on a directed graph $\zug{V, E}$, where $V = V_1 \cup V_2$ is a fixed partition of the vertices. 
The game proceeds as follows. A token is initially placed on some vertex. When the token is placed on $v \in V_i$, for $i \in \set{1,2}$, \PLi chooses $u$ with $\zug{v, u} \in E$ to move the token to.
The outcome of the game is an infinite path, called a {\em play}. We focus on {\em reachability} games: \PO wins a play iff it visits a set of target vertices $T \subseteq V$.

In this paper, we introduce {\em pawn games}, which are graph games in which the control of vertices changes dynamically throughout the game as follows. 
The arena consists of $d$ {\em pawns}. For $1 \leq j \leq d$, Pawn~$j$ {\em owns} a set of vertices $V_j$. Throughout the game, the pawns are distributed between the two players, and in each turn, the control of pawns determines which player moves the token. Pawn control may be updated after moving the token by running a predetermined {\em mechanism}. 
Formally, a {\em configuration} of a pawn game is a pair $\zug{v, P}$, where $v$ denotes the position of the token and $P$ the set of pawns that \PO controls. The player who moves the token is determined according to $P$: if \PO controls a pawn that owns $v$, then \PO moves. Specifically, when each vertex is owned by a unique pawn, i.e., $V_1,\ldots, V_d$ partitions $V$, then \PO moves iff he controls the pawn that owns $v$. 
We consider the following mechanisms for exchanging control of pawns. 
For $i \in \set{1,2}$, we denote by $\ni= 3-i$ the ``other player''.

\begin{itemize}
\item{\bf Optional grabbing.} For $i \in \set{1,2}$, following a \PLi move, \PLni has the \emph{option} to {\em grab} one of \PLi's pawns; namely, transfer one of the pawns that \PLni to his control. 
\item{\bf Always grabbing.} For $i \in \set{1,2}$, following \emph{every} \PLi move, \PLni grabs one of \PLi's pawns. 
\item{\bf Always grabbing or giving.} Following a \PLi move, \PLni either grabs one of \PLi's pawns or gives her one of his pawns.
\item{\bf $k$-grabbing.} For $k \in \Nat$, \PO can grab at most $k$ pawns from \PT throughout the game. In each round, after moving the token, \PO has the option of grabbing one of the pawns that is controlled by \PT.  A grabbed pawn stays in the control of \PO for the remainder of the game. Note the asymmetry: only \PO grabs pawns.
\end{itemize}

Note that players in pawn games have two types of actions: moving the token and transferring control of pawns. 
We illustrate the model and some interesting properties of it. 

\begin{exa}
\label{ex:detrimental}
Consider the game $\G_1$ in Fig.~\ref{fig:example}(left). We consider optional-grabbing and the same reasoning applies for always-grabbing.  
Each vertex is owned by a unique pawn, and \PO's target is $t$. Note that \PT wins if the game reaches $s$. We claim that $\G_1$ is {\em non-monotonic}: increasing the set of pawns that \PO initially controls is ``harmful'' for him. Formally, \PO wins from configuration $\zug{v_0, \emptyset}$, i.e., when he initially does not control any pawns,
but loses from $\zug{v_0, \set{v_0}}$, i.e., when controlling $v_0$. Indeed, from $\zug{v_0, \emptyset}$, \PT initially moves the token from $v_0$ to $v_1$, \PO then uses his option to grab $v_1$, and wins by proceeding to $t$. Second, from $\zug{v_0, \set{v_0}}$, \PO makes the first move and thus cannot grab $v_1$. Since \PT controls $v_1$, she wins by proceeding to $s$. 
In Thm.~\ref{thm:optional-grabbing-detrimental} and~\ref{thm:always-grabbing-detrimental}, we generalize this observation and show, somewhat surprisingly, that if a player wins from the current vertex $v$, then he wins from $v$ with fewer pawns as long as if he controlled $v$ previously, then he maintains control of $v$. 

Consider the game $\G_2$ in Fig.~\ref{fig:example} (right). We consider optional-grabbing, each vertex is owned by a unique pawn, and \PO's target is $t$. We claim that \PO wins from configuration $\zug{v_0, \set{v_0, v_2}}$ and \PT can force the game to visit $v_1$ twice. This differs from turn-based games in which if \PO wins, he can force winning while visiting each vertex at most once. To illustrate, consider the following outcome. \PO makes the first move, so he cannot grab $v_1$. \PT avoids losing by moving to $v_2$. \PO will not grab, move to $v_3$, \PT moves to $v_1$, then \PO grabs $v_1$ and proceeds to $t$.
We point out that no loop is closed in the explicit {\em configuration graph} that corresponds to $\G_2$.

%following is a of code that helps draw square nodes in figures.
\makeatletter
\def\squarecorner#1{
    \pgf@x=\the\wd\pgfnodeparttextbox%
    \pgfmathsetlength\pgf@xc{\pgfkeysvalueof{/pgf/inner xsep}}%
    \advance\pgf@x by 2\pgf@xc%
    \pgfmathsetlength\pgf@xb{\pgfkeysvalueof{/pgf/minimum width}}%
    \ifdim\pgf@x<\pgf@xb%
        \pgf@x=\pgf@xb%
    \fi%
    \pgf@y=\ht\pgfnodeparttextbox%
    \advance\pgf@y by\dp\pgfnodeparttextbox%
    \pgfmathsetlength\pgf@yc{\pgfkeysvalueof{/pgf/inner ysep}}%
    \advance\pgf@y by 2\pgf@yc%
    \pgfmathsetlength\pgf@yb{\pgfkeysvalueof{/pgf/minimum height}}%
    \ifdim\pgf@y<\pgf@yb%
        \pgf@y=\pgf@yb%
    \fi%
    \ifdim\pgf@x<\pgf@y%
        \pgf@x=\pgf@y%
    \else
        \pgf@y=\pgf@x%
    \fi
    \pgf@x=#1.5\pgf@x%
    \advance\pgf@x by.5\wd\pgfnodeparttextbox%
    \pgfmathsetlength\pgf@xa{\pgfkeysvalueof{/pgf/outer xsep}}%
    \advance\pgf@x by#1\pgf@xa%
    \pgf@y=#1.5\pgf@y%
    \advance\pgf@y by-.5\dp\pgfnodeparttextbox%
    \advance\pgf@y by.5\ht\pgfnodeparttextbox%
    \pgfmathsetlength\pgf@ya{\pgfkeysvalueof{/pgf/outer ysep}}%
    \advance\pgf@y by#1\pgf@ya%
}
\makeatother
\pgfdeclareshape{square}{
    \savedanchor\northeast{\squarecorner{}}
    \savedanchor\southwest{\squarecorner{-}}

    \foreach \x in {east,west} \foreach \y in {north,mid,base,south} {
        \inheritanchor[from=rectangle]{\y\space\x}
    }
    \foreach \x in {east,west,north,mid,base,south,center,text} {
        \inheritanchor[from=rectangle]{\x}
    }
    \inheritanchorborder[from=rectangle]
    \inheritbackgroundpath[from=rectangle]
}

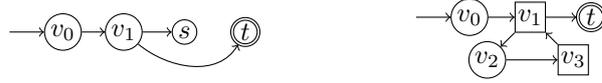
\begin{figure}[t]
\begin{minipage}[b]{0.49\linewidth}
\centering
\begin{tikzpicture}[node distance={8mm},initial text = ,main/.style = {draw, circle, text=black, inner sep=1pt}] 
\node[main,initial] (1) {$v_0$};
\node[main] (2) [right of=1] {$v_1$};
\node[main] (3) [right of=2] {$s$};
\node[main,accepting] (4) [right of=3] {$t$};
\draw[->] (1) -- (2);
\draw[->] (2) -- (3);
\draw[->] (2) to [out=320,in=240,looseness=1] (4);
\end{tikzpicture}
\end{minipage}
\begin{minipage}[b]{0.5\linewidth}
\begin{tikzpicture}[node distance={8mm},initial text = ,main/.style = {draw, circle, text=black, inner sep=1pt},gain/.style = {draw, square, text=black, inner sep=1pt}] 
\node[main,initial] (1) {$v_0$};
\node[gain] (2) [right of=1] {$v_1$};
\node[main,accepting] (3) [right of=2] {$t$};
\node[main] (5) [below left of=2] {$v_2$};
\node[gain] (6) [below right of=2] {$v_3$};
\draw[->] (1) -- (2);
\draw[->] (2) -- (5);
\draw[->] (5) -- (6);
\draw[->] (6) -- (2);
\draw[->] (2) -- (3);
\end{tikzpicture}
\end{minipage}
\caption{Left: The pawn game $\G_1$; a non-monotonic game under optional-grabbing. Right: The pawn game $\G_2$ in which \PO wins from $\zug{v_0, \set{v_0, v_1}}$, but must visit $v_1$ twice.}
\label{fig:example}
\end{figure}
\end{exa}

\paragraph*{Applications}
Pawn games model multi-agent settings in which the agent who acts in each turn is not predetermined. We argue that such settings arise naturally. 

\noindent{\bf Quantitative shield synthesis.} 
It is common practice to model an {\em environment} as a Kripke structure (e.g.~\cite{SST18}), which for sake of simplicity, we will think of as a graph in which vertices model environment states and edges model actions. A {\em policy} chooses an outgoing edge from each vertex. A popular technique to obtain policies is {\em reinforcement learning} (RL)~\cite{SB98} whose main drawback is lack of worst-case guarantees~\cite{BCGPR21}. In order to regain safety at runtime, a {\em shield}~\cite{KAB+17,AB+19,BCGPR21} is placed as a proxy: in each point in time, it can alter the action of a policy. The goal in {\em shield synthesis} is to synthesize a shield {\em offline} that ensures safety at runtime while minimizing interventions. We suggest a procedure to synthesize shields based on $k$-grabbing pawn games. \PT models an unknown policy.
We set his goal to reaching an unsafe state. 
\PO (the shield) ensures safety by grabbing at most $k$ times. Grabbing is associated with a shield intervention. Note that once the shield intervenes in a vertex $v$, it will choose the action at $v$ in subsequent turns. 
An optimal shield is obtained by finding the minimal $k$ for which \PO has a winning strategy.

We describe other examples that can be captured by a $k$-grabbing pawn game in which, as in the shield application, \PO models an ``authority'' that has the ``upper hand'', and aims to maximize freedom of action for \PT while using grabs to ensure safety. Consider a concurrent system in which \PT models a scheduler and \PO can force synchronization, e.g., by means of ``locks'' or ``fences'' in order to maintain correctness (see~\cite{CC+17}). Synchronization is minimized in order to maximize parallelism and speed. As another example, \PO might model an operating system that allows freedom to an application and blocks only unsafe actions.
As a final example, in~\cite{AA+20}, synthesis for a safety specification was enriched with ``advice'' given by an external policy for optimizing a soft quantitative objective. Again, the challenge is how to maximize accepting advice while maintaining safety. 
\noindent{\bf Modelling crashes.}
A {\em sabotage game}~\cite{Ben05} is a two-player game which is played on a graph. \PO (the Runner) moves a token throughout the graph with the goal of reaching a target set. In each round, \PT (the Saboteur) crashes an edge from the graph with the goal of preventing \PO from reaching his target.
Crashes are a simple type of fault that restrict \PO's actions. A malicious fault (called {\em byzantine faults}~\cite{LSP82}) actively tries to harm the network, e.g., by moving away from the target.
Pawn games can model sabotage games with byzantine faults: each vertex (router) is owned by a unique pawn, all pawns are initially owned by \PO, and a \PT grab corresponds to a byzantine fault.
Several grabbing mechanisms are appealing in this context: $k$-grabbing restricts the number of faults and optional- and always-grabbing accommodate  repairs of routers.

\paragraph*{Our results}
We distinguish between three types of ownership of vertices. Let $V = V_1 \cup \ldots \cup V_d$ be a set of vertices, where for $j \in \set{1,\ldots, d}$, Pawn~$j$ owns the vertices in $V_j$. 
In {\em one vertex per pawn} (OVPP) games, each pawn owns exactly one vertex, thus $V_j$ is a singleton, for all $j \in \set{1,\ldots, d}$. In {\em multiple vertices per pawn} (MVPP) games, $V_1,\ldots,V_d$ consists of a partition of $V$, where the sets might contain more than one vertex. In {\em overlapping multiple vertices per pawn} (OMVPP) games, the sets might overlap. For example, in the shield synthesis application above, the type of ownership translates to dependencies between interventions: OVPP models no dependencies, MVPP models cases in which interventions come in ``batches'', e.g., grabbing control in all states labeled by some predicate, and OMVPP models the case when the batches overlap. 
We define that \PO moves the token from a vertex $v$ iff he controls at least one of the pawns that owns $v$. Clearly, OMVPP generalizes MVPP, which in turn generalizes OVPP. 

We consider the problem of deciding whether \PO wins a reachability pawn game from an initial configuration of the game. Our results are summarized below. 

\begin{table}[ht!]
\centering
\resizebox{\textwidth}{!}{
\begin{tabular}{|l|l|l|l|}
\hline
Mechanisms & OVPP & MVPP & OMVPP\\
\hline
$k$-grabbing & \begin{tabular}[c]{@{}l@{}}PTIME (Thm. \ref{thm:OVPP-k-grabbing}) 
\end{tabular} & NP-hard (Thm. \ref{thm:reach-MVPP-k-grabbing}) & PSPACE-C (Thm. \ref{thm:PSPACE}) \\
\hline
Optional-grabbing & PTIME (Thm. \ref{thm:optional-grabbing-PTIME}) & EXPTIME-C (Thm. \ref{thm-MVPP optional grab}) & EXPTIME-C (Thm. \ref{thm-MVPP optional grab}) \\
\hline
Always & \begin{tabular}[c]{@{}l@{}}PTIME \\  (grab or give; Thm.~\ref{thm:always-grab-or-give}) \end{tabular} & \begin{tabular}[c]{@{}l@{}}PTIME (grab or give; Thm.~\ref{thm:always-grab-or-give}) \\ EXPTIME-C (grab; Thm. \ref{thm-MVPP always grab})\end{tabular} & EXPTIME-C (grab; Thm. \ref{thm-MVPP always grab})\\                                        
\hline
\end{tabular}}
\end{table}

Pawn games are succinctly-represented turn-based games. A naive algorithm to solve a pawn game constructs and solves an explicit turn-based game on its configuration graph leading to membership in $\EXP$.
We thus find the positive results to be pleasantly surprising; we identify classes of succinctly-represented games that can be solved in PTIME. Each of these algorithms is obtained by a careful and tailored modification to the {\em attractor-computation} algorithm for turn-based reachability games. For OMVPP $k$-grabbing, the PSPACE upper bound is obtained by observing that grabs in a winning strategy must be spaced by at most $|V|$ turns, implying that a game ends within polynomial-many rounds (Lem.~\ref{lem:stepsOMVPP}).

Our $\EXP$-hardness proofs are based on a new class of games called {\em \LAK} games and may be of independent interest. A \LAK game is a turn-based game that is enriched with a set of locks, where each lock is associated with a key. 
Each edge is labeled by a subset of locks and keys. A lock can either be {\em closed} or {\em open}. An edge that is labeled with a closed lock cannot be crossed. A lock changes state once an edge labeled by its key is traversed.
We show two reductions. The first shows that deciding the winner in \LAK games is $\EXP$-hardness. Second, we reduce \LAK games to MVPP optional-grabbing pawn games. The core of the reduction consists of gadgets that simulate the operation of locks and keys using pawns. Then, we carefully analyze the pawn games that result from applying both reductions one after the other, and show that the guarantees are maintained when using always grabbing instead of optional grabbing. 
The main difficulty in constructing a winning \PLi strategy under always-grabbing from a winning \PLi strategy under optional-grabbing is to ensure that throughout the game, both players have \emph{sufficient} and the \emph{correct} pawns to grab (Lem.~\ref{lem:always-grab_optional-grab}).

\paragraph*{Related work} 
The semantics of pawn games is inspired by the seminal paper~\cite{AHK02}. 
There, the goal is, given a game, an objective $O$, and a set $C$ of pawns (called ``players'' there), to decide whether \PO (called a ``coalition'' there) can ensure $O$ when he controls the pawns in $C$. 
A key distinction from pawn games is that the set $C$ that is controlled by \PO is fixed. The paper introduced 
a logic called {\em alternating time temporal logic}, which was later significantly extended and generalized to {\em strategy logic} \cite{CHP10,MMPV14,MMV10}. 
%, first in two-player games~\cite{CHP10} and later in multi-player games, e.g., \cite{MMPV14,MMV10}. 
Multi-player games with rational players have been widely studied; e.g., finding Nash equilibrium~\cite{Ume11} or subgame perfect equilibrium~\cite{BRB21}, and rational synthesis~\cite{FKL10,KPV16,WG+16,BC+16}. 
A key distinction from pawn games is that, in pawn games, as the name suggests, the owners of the resources (pawns) have no individual goals and act as pawns in the control of the players. Changes to multi-player graph games in order to guarantee existence or improve the quality of an equilibrium
have been studied~\cite{AAK15,Per19,BK20,KS23}. The key difference from our approach is that there, changes occur {\em offline}, before the game starts, whereas in pawn games, the transfer of vertex ownership occurs {\em online}.
In {\em bidding games}~\cite{LLPSU99,AHC19} (see in particular, {\em discrete-bidding games}~\cite{DP10,AAH21,AS25}) control of vertices changes online: players have budgets, and in each turn, a {\em bidding} determines which player moves the token.  
Bidding games are technically very different from pawn games. While pawn games allow varied and fine-grained mechanisms for transfer of control, bidding games only consider strict auction-based mechanisms, which lead to specialized proof techniques that cannot be applied to pawn games. For example, bidding games are monotonic -- more budget cannot harm a player -- whereas pawn games are not (see Ex.~\ref{ex:detrimental}).

\section{Preliminaries}

For $k \in \Nat$, we use $[k]$ to denote the set $\set{1, \dots, k}$. 
For $i \in \set{1,2}$, we use $\ni = 3-i$ to refer to the ``other player''. 

\paragraph*{Turn-based games}
Throughout this paper we consider {\em reachability} objectives. For general graph games, see for example~\cite{AG11}.
A {\em turn-based game} is $\G = \zug{V, E, T}$, where $V = V_1 \cup V_2$ is a set of vertices that is partitioned among the players, $E \subseteq V \times V$ is a set of directed edges, and $T \subseteq V$ is a set of target vertices for \PO. \PO's goal is to reach $T$ and \PT's goal is to avoid $T$. For $v \in V$, we denote the {\em neighbors} of $v$ by $N(v) = \set{u \in V: E(v,u)}$.
Intuitively, a {\em strategy} is a recipe for playing a game: in each vertex it prescribes a neighbor to move the token to. 
Formally, for $i \in \set{1, 2}$, a strategy for \PLi is a function $f: V_i \rightarrow V$ such that for every $v \in V_i$, we have $f(v) \in N(v)$.\footnote{We restrict to {\em memoryless} strategies since these suffice for reachability objectives.}
An initial vertex $v_0 \in V$ together with two strategies $f_1$ and $f_2$ for the players, give rise to a unique {\em play}, denoted $\pi(v_0,f_1,f_2)$, which is a finite or infinite path in $\G$ and is defined inductively as follows. The first vertex is $v_0$. For $j \geq 0$, assuming $v_0,\ldots, v_j$ has been defined, then $v_{j+1} = f_i(v_j)$, where $v_j \in V_i$, for $i \in \set{1,2}$. A \PO strategy $f_1$ is \emph{winning} from $v_0 \in V$ if for every \PT strategy $f_2$, the play $\play(v_0, f_1, f_2)$ ends in $T$. Dually, a \PT strategy $f_2$ is winning from $v_0 \in V$ if for every \PO strategy $f_1$, the play $\play(v_0, f_1, f_2)$ does not visit $T$. 

\begin{thmC}[\cite{GTW02}]
\label{thm:turn-based}
 Turn based games are {\em determined}: from each vertex, one of the players has a (memoryless) winning strategy.  Deciding the winner of a game is in PTIME.
\end{thmC}
\begin{proof}[Proof sketch]
For completeness, we briefly describe the classic {\em attractor-computation} algorithm. Consider a game $\zug{V, E, T}$. Let $W_0 = T$. For $i \geq 1$, let $W_i = W_{i-1} \cup \set{v \in V_1: N(v) \cap W_i \neq \emptyset} \cup \set{v \in V_2: N(v) \subseteq W_i}$. One can prove by induction that $W_i$ consists of the vertices from which \PO can force reaching $T$ within $i$ turns. The sequence necessarily reaches a {\em fixed point} $W^1 = \bigcup_{i \geq 1} W_i$, which can be computed in linear time. Finally, one can show that \PT has a winning strategy from each $v \notin W^1$.
\end{proof}

\paragraph*{Pawn games} 
A {\em pawn game} with $d \in \Nat$ pawns is 
$\P = \zug{V, E, T, \M}$, 
where $V = V_1 \cup \ldots \cup V_d$ and for $j \in [d]$, $V_j$ denotes the vertices that Pawn~$j$ {\em owns}, $E$ and $T$ are as in turn-based games, 
and $\M$ is a mechanism for exchanging pawns as we elaborate later. 
\PO wins a play if it reaches $T$. 
We stress that  the set of pawns that he controls when reaching $T$ is irrelevant. We omit $\M$ when it is clear from the context.
We distinguish between classes of pawn games based on the type of ownership of vertices:

\begin{itemize}
\item{\bf One Vertex Per Pawn (OVPP).} There is a one-to-one correspondence between pawns and vertices; namely, $|V| = d$ and each $V_j$ is singleton, for $j \in [d]$. For $j \in [d]$ and $\set{v_j} = V_j$, we sometimes abuse notation by referring to Pawn~$j$ as $v_j$.
\item{\bf Multiple Vertices Per Pawn (MVPP).} Each vertex is owned by a unique pawn but a pawn can own multiple vertices, thus $V_1, \ldots, V_d$ is a partition of $V$.
\item{\bf Overlapping Multiple Vertices Per Pawn (OMVPP).} Each pawn can own multiple vertices and a vertex can be owned by multiple pawns, i.e., we allow $V_i \cap V_j \neq \emptyset$, for $i \neq j$.
\end{itemize}
Clearly OMVPP generalizes MVPP, which generalizes OVPP.
In MVPP too, we sometimes abuse notation and refer to a pawn by a vertex that it owns.

A {\em configuration} of a pawn game is $\zug{v, P}$, meaning that the token is placed on a vertex $v \in V$ and $P \subseteq [d]$ is the set of pawns that \PO~{\em controls}. Implicitly, \PT controls the complement set $\overline{P}= [d] \setminus P$. \PO moves the token from $\zug{v, P}$ iff he controls at least one pawn that owns $v$. Note that in OVPP and MVPP, let $j \in [d]$ with $v \in V_j$, then \PO moves iff $i \in P$. Once the token moves, we update the control of the pawns by applying $\M$. 

\paragraph*{From pawn games to turn-based games} 
We describe the formal semantics of pawn games together with the pawn-exchanging mechanisms by describing the explicit turn-based game that corresponds to a pawn game. For a pawn game $\G = \zug{V, E, T, \M}$, we construct the turn-based game $\G' = \zug{V', E', T'}$. For $i \in \set{1,2}$, denote by $V'_i$ \PLi's vertices in $\G'$. 
The vertices of $\G'$ consist of two types of vertices: configuration vertices $C = V \times 2^{[d]}$, and {\em intermediate} vertices $V \times C$. When $\M$ is $k$-grabbing, configuration vertices include the remaining number of pawns that \PO can grab, as we elaborate below. The target vertices are $T' = \set{\zug{v, P}: v \in T}$. We describe $E'$ next.
For a configuration vertex $c = \zug{v, P}$, we define $c \in V'_1$ iff there exists $j \in P$ such that $v \in V_j$. That is, \PO moves from $c$ in $\G'$ iff he moves from $c$ in $\G$. We define the neighbors of $c$ to be the intermediate vertices $\set{\zug{v', c}: v' \in N(v)}$. That is, moving the token in $\G'$ from $c$ to $\zug{v', c}$ corresponds to moving the token from $v$ to $v'$ in $\G$. Moves from intermediate vertices represent an application of $\M$. We consider the following mechanisms.

\noindent{\bf Optional grabbing.} For $i \in \set{1,2}$, following a \PLi move, \PLni has the option to grab one of \PLi's pawns. %, that is, a player a may or may not grab a pawn from the other player following a move of the other player.
Formally, for a configuration vertex $c = \zug{v, P} \in V'_1$, we have $N(c) \subseteq V'_2$. From $\zug{v', c} \in N(c)$, \PT has two options: (1) do not grab and proceed to $\zug{v', P}$, or (2) grab $j \in P$, and proceed to $\zug{v', P \setminus \set{j}}$. The definition for \PT is dual.

\noindent{\bf Always grabbing.} For $i \in \set{1,2}$, following a \PLi move, \PLni always has to grab one of \PLi's pawns. The formal definition is similar to optional grabbing with the difference that option (1) of not grabbing is not available to the players. We point out that \PLni grabs only after \PLi has moved, which in particular implies that \PLi controls at least one pawn that \PLni can grab.

\noindent{\bf Always grabbing or giving.} Following a \PLi move, \PLni must either grab one of \PLi's pawns or {\em give} her a pawn. The formal definition is similar to always grabbing with the difference that, for an intermediate vertex $\zug{v', \zug{v, P}}$, there are both neighbors of the form $\zug{v', P \setminus \set{j}}$, for $j \in P$, and neighbors of the form $\zug{v', P \cup \set{j}}$, for $j \notin P$.

\noindent{\bf $k$-grabbing.} After each round, \PO has the option of grabbing a pawn from \PT, and at most $k$ grabs are allowed in a play. 
A configuration vertex in $k$-grabbing is $c = \zug{v, P, r}$, where $r \in [k] \cup \set{0}$ denotes the number of grabs remaining. 
Intermediate vertices are \PO vertices. Let $\zug{v', c} \in V'_1$. 
\PO has two options: (1) do not grab and proceed to the configuration vertex $\zug{v', P, r}$, or (2) grab $j \notin P$, and proceed to $\zug{v', P \cup \set{j}, r-1}$ when $r > 0$.
Note that grabs are not allowed when $r=0$ and that Pawn~$j$ stays at the control of \PO for the remainder of the game.

Since pawn games are succinctly-represented turn-based games, Thm.~\ref{thm:turn-based} implies determinacy; namely, one of the players wins from each initial configuration. We study the problem of determining the winner of a pawn game, formally defined as follows. 

\begin{defi}
Let $\alpha \in \set{\text{OVPP, MVPP, OMVPP}}$ and $\beta$ be a pawn-grabbing mechanism.
The problem $\alpha$ $\beta$ PAWN-GAMES takes as input an $\alpha$ $\beta$ pawn game $\G$ and an initial configuration $c$, and the goal is to decide whether \PO wins from $c$ in $\G$.
\end{defi}

A naive algorithm to solve a pawn game applies attractor computation on the explicit turn-based game, which implies the following theorem. 

\begin{thm}
\label{thm:exp}
$\alpha$ $\beta$ PAWN-GAMES is in EXPTIME, for all values of $\alpha$ and $\beta$.
\end{thm}

%%%%%%%%%%%%%%%%%%%%%%%%%%%%%%%%%%%%%%%%%%%%%%%%%%%%%%%%%%

\section{Optional-Grabbing Pawn Games}\label{sec:Optional-Grabbing Pawn Games}
Before describing our complexity results, we identify a somewhat unexpected property of MVPP optional-grabbing games.
Consider a vertex $v$ and two sets of pawns $P$ and $P'$ having $P' \subseteq P$. Intuitively, it is tempting to believe that \PO ``prefers'' configuration $c = \zug{v, P}$ over $c' = \zug{v, P'}$ since he controls more pawns in $c$. Somewhat surprisingly, the following theorem shows that the reverse holds (see also Ex.~\ref{ex:detrimental}). More formally, the theorem states that if \PO wins from $c$, then he also wins from $c'$, under the restriction that if he makes the first move at $c$ (i.e., he controls $v$ in $c$), then he also makes the first move in $c'$ (i.e., he controls $v$ in $c'$).

\begin{thm}
\label{thm:optional-grabbing-detrimental}  
 Consider a configuration $\zug{v, P}$ of an MVPP optional-grabbing pawn game $\G$. Let $j \in [d]$ such that $v \in V_j$ and $P' \subseteq P$. Assuming that $j \in P$ implies $j \in P'$, if \PO wins from $\zug{v, P}$, he wins from $\zug{v, P'}$. Assuming that $j \in \overline{P'}$ implies $j \in \overline{P}$, if \PT wins from $\zug{v, P'}$, she wins from $\zug{v, P}$. 
\end{thm}
\begin{proof}
We prove for \PO and the proof for \PT follows from determinacy. 
Let $\G$, $P$, $P'$, $c = \zug{v, P}$, and $c' = \zug{v, P'}$ be as in the theorem statement.
Let $\G'$ be the turn-based game corresponding to $\G$. For $i \geq 0$, let $W_i$ be the set of vertices in $\G'$ from which \PO can win in at most $i$ rounds (see Thm.~\ref{thm:turn-based}). 
The following claim clearly implies the theorem. 
Its proof proceeds by a careful induction.

\noindent{\bf Claim:} Configuration vertices: for $i \geq 0$, if $\zug{v,P}\in W_i$, then $\zug{v,P'}\in W_i$.
Intermediate vertices: for $i \geq 1$ and every vertex $u \in N(v)$, if $\zug{u,c}\in W_{i-1}$, then $\zug{u,c'}\in W_{i-1}$.

For the base case, we prove for both $i=0$ and $i=1$. Recall that the target set of $\G'$, which coincides with $W_0$, consists of configuration vertices whose $V$ component is in $T$. Thus, for the first part of the claim, since $c \in W_0$, then $v \in T$, and thus $\zug{v, P'} \in W_0$. Recall that an intermediate vertex is of the form $\zug{u, b}$, where $u$ denotes the ``next'' location of the token and $b$ denotes the ``previous'' configuration. In the case of $W_1$, since \PO wins in one turn, each vertex in $W_1$ is of the form $\zug{u, b}$, where $u \in T$. Thus, for $\zug{u,c}\in W_1$, we clearly have $\zug{u,c'}\in W_1$.

For the induction hypothesis, assume that the claim holds for $i=n$, and we prove for $i=n+1$.
We start with configuration vertices.
Assume $c=\zug{v,P}\in W_{n+1}$. 
We will show that $c'=\zug{v,P'}\in W_{n+1}$.
Recall that $N(c)$ consist of intermediate vertices of the form $\zug{u, c}$, where $u \in N(v)$, thus $\zug{u, c} \in N(c)$ iff $\zug{u, c'} \in N(c')$. Note that if $\zug{u, c} \in N(c)$ is winning for \PO, i.e., $\zug{u, c} \in W_n$, then by the induction hypothesis, $\zug{u, c'} \in W_n$. We distinguish between two cases. 
In the first case, \PO controls $c$, i.e., $j\in P$ and $c$ is a \PO vertex. In this case, our assumption is that $c'$ is also a \PO vertex. Since $c \in W_{n+1}$, there must exist a neighbor $\zug{u, c}$ of $c$ having $\zug{u,c} \in W_n$. By the above, $\zug{u, c'} \in (N(c') \cap W_n)$, thus $c' \in W_{n+1}$. In the second case, \PT controls $c$, thus $j\notin P$. Recall that $P' \subseteq P$, thus \PT controls $c'$ as well. Note that $c \in W_{n+1}$ implies that \PO wins from all of its neighbors
%shibashis-postcr
of the form $\zug{u,c}$.
Now consider the contrapositive of the assumption that states that $j \notin P'$ implies that $j \notin P$ which holds here and since $\zug{u,c} \in W_n$, by induction hypothesis, it follows that $\zug{u,c'} \in W_n$.
Thus, \PO wins from all the neighbors of $c'$, thus $c' \in W_{n+1}$.

We turn to the second part of the claim that addresses intermediate vertices. 
We again distinguish between two cases: $j \in P$ and $j \notin P$. 
Consider an intermediate vertex $\zug{u, c} \in W_{n+1}$.
We will show that $\zug{u, c'} \in W_{n+1}$.
We denote by $\ell$, the pawn that owns $u$. 

%shibashis-postcr
First consider the case $j \in P$.
Recall that in optional grabbing, \PT has the option of grabbing after \PO moves, thus $\zug{u, c}$ is a \PT vertex. 
Since the claim requires that if $j \in P$, then $j \in P'$, we conclude that $\zug{u, c'}$ is also a \PT vertex. 
Consider a neighbor of $\zug{u, c'}$, a configuration vertex $\zug{u,Q'}$. 
Note that either \PT does not use the option to grab, then $Q'=P'$, or she grabs a pawn $r$ from \PO, then $Q'=P' \setminus \set{r}$.
Note that in order to apply the induction hypothesis on $\zug{u, Q'}$, we need to find a neighbor $\zug{u, Q}$ of $\zug{u, c}$ such that if $\ell \notin Q'$, then $\ell \notin Q$. 
Note that at $\zug{u, c}$, \PT has the option of not grabbing as well as of grabbing $\ell$, thus both $\zug{u, P}$ and $\zug{u, P\setminus \set{\ell}}$ are neighbors of $\zug{u, c}$. Note that $P = P \setminus \set{\ell}$ when $\ell \notin P$. That is, \PT cannot grab $\ell$ when she already owns it. Since $\zug{u, c} \in W_{n+1}$ and it is a \PT vertex, both $\zug{u, P} \in W_n$ and $\zug{u, P\setminus \set{\ell}} 
\in W_n$. 
Finally, if $\ell \in Q'$, define $Q := P$, and if $\ell \notin Q'$, define $Q := P \setminus \set{\ell}$. 
In both cases, since $Q' \subseteq P'$ and $P' \subseteq P$, we have $Q' \subseteq Q$ and meets the requirement in the claim on $\ell$. 
Thus, since $\zug{u, Q} \in W_n$, by the induction hypothesis, $\zug{u, Q'} \in W_n$.
Since $\zug{u, Q'}$ is any arbitrary neighbour of the \PT vertex $\zug{u,c'}$, we have that $\zug{u,c'} \in W_{n+1}$.

In the second case $j\notin P$. That is, \PT makes the move in $\zug{v, P}$, and thus $\zug{u, c}$ is a \PO vertex. Since $P' \subseteq P$, we have $j \notin P'$, thus $\zug{u, c'}$ is also a \PO vertex. Since $\zug{u, c} \in W_{n+1}$, it has a neighbor $\zug{u, Q} \in W_n$. 
Note that since \PO either grabs or does not grab in $\zug{u, c}$, we have $P \subseteq Q$. We find a neighbor $\zug{u, Q'}$ of $\zug{u, c'}$ to apply the induction hypothesis on.
Note that \PO has the option to grab at $\zug{u, c'}$, and, among his choices, he can choose not to grab or to grab $\ell$.
If $\ell \in Q$, then we choose $Q' := P' \cup \set{\ell}$, and if $\ell \notin Q$, we choose $Q' := P'$. In both cases, $Q' \subseteq Q$ and meets the requirement in the claim. 
Thus, since $\zug{u,Q} \in W_n$, by the induction hypothesis, we have $\zug{u, Q'} \in W_n$, and hence $\zug{u,c'} \in W_{n+1}$.
\end{proof}

The following corollary of Thm.~\ref{thm:optional-grabbing-detrimental} shows that we can restrict attention to ``locally-grabbing'' strategies that only grab the pawn that owns the vertex on which the token is placed. In other words, a locally-grabbing strategy will not grab a pawn if the token is not placed on the vertex that it owns. 

\begin{cor}\label{cor:optional-grabbing-detrimental}
Consider an MVPP optional-grabbing game. Suppose that \PO controls $P \subseteq [d]$, and that \PT moves the token to a vertex $v$ owned by Pawn~$j$, i.e., $v \in V_j$. \PO has the option to grab. If \PO can win by grabbing a pawn $j' \neq j$, i.e., a pawn that does not own the next vertex, he can win by not grabbing at all. Formally, if \PO wins from $\zug{v, P \cup \set{j'}}$, he also wins from $\zug{v, P}$. And dually for \PT. 
\end{cor}
%shibashis-cr
We point out that Thm.~\ref{thm:optional-grabbing-detrimental} and Cor.~\ref{cor:optional-grabbing-detrimental} do not hold for OMVPP optional-grabbing games.

\subsection{OVPP: A PTIME algorithm}

We turn to study complexity results, and start with the following positive result.

\begin{thm}
\label{thm:optional-grabbing-PTIME}
OVPP optional-grabbing PAWN-GAMES is in PTIME.
\end{thm}
\begin{proof}
We describe the intuition of the algorithm, the pseudo-code can be found in Alg.~\ref{algo:OVPP_optional}.
Recall that in turn-based games (see Thm.~\ref{thm:turn-based}), the attractor computation iteratively ``grows'' the set of states from which \PO wins: initially $W_0=T$, and in each iteration, a vertex $u$ is added to $W_i$ if (1) $u$ belongs to \PT and all its neighbors belong to $W_i$ or (2) $u$ belongs to \PO and it has a neighbor in $W_i$. 
In optional-grabbing games, applying attractor computation is intricate since vertex ownership is dynamic. 
Note that the reasoning behind (1) above holds; namely, if $N(u) \subseteq W_i$, no matter who controls $u$, necessarily $W_i$ is reached in the next turn. 
However, the reasoning behind (2) fails. 
Consider a \PO vertex $u$ that has two neighbors $v_1 \in W_i$ and $v_2 \notin W_i$. While $u$ would be in $W_{i+1}$ according to (2), under optional-grabbing, when \PO makes the move into $u$, \PT can avoid $W_i$ by grabbing $u$ and proceeding to $v_2$. 

In order to overcome this, our algorithm operates as follows. Vertices that satisfy (1) are added independent of their owner (Line~$4$). The counterpart of (2) can be seen as two careful steps of attractor computation. First, let $B$ denote the {\em border} of $W_i$, namely the vertices who have a neighbor in $W_i$ (Line~$6$). 
Second, a vertex $u$ is in $W_{i+1}$ in one of two cases. (i) $u \in B$ and all of its neighbors are in $B \cup W_i$ (Line~$10$). Indeed, if \PO controls $u$ he wins by proceeding to $W_i$ and if \PT owns $u$, she can avoid $W_i$ by moving to $B$, then \PO grabs and proceeds to $W_i$. (ii) \PT controls $u$ in the initial configuration and all of its neighbors are in $B$ (Line~$12$). Indeed, \PT cannot avoid proceeding into $B$, and following \PT's move, \PO grabs and proceeds to $W_i$.

More formally, consider an input OVPP optional-grabbing game $\G = \zug{V, E, T}$ and an initial configuration $\zug{v_0,P_0}$. Correctness of the algorithm follows from the two following claims. First, we show soundness; namely, whenever the algorithm returns that \PO wins, he does indeed have a winning strategy from $\zug{v_0, P_0}$. 

\begin{clm}
For $i \geq 0$, \PO wins from every vertex $v \in W_i$ no matter who makes the last step into $v$.     
\end{clm}
\begin{proof}
The proof is by induction on $i$. The base case is trivial since $W_0 = T$. For the inductive step, suppose that the claim holds for $W_i$ and we show that it holds in all ways that it can be extended. 

\noindent\underline{Case 1 -- Line~$4$:} for a vertex $u$ having $N(u) \subseteq W_i$, \PO wins no matter who controls $u$ since in the next turn, $W_i$ is necessarily reached. 

\noindent\underline{Case 2 -- Line~$10$:} We claim that \PO wins from $u \in B'$ no matter who controls $u$. Indeed, if \PO controls $u$, he wins by proceeding to $W_i$. If \PT controls $u$ and avoids $W_i$, then the next vertex is necessarily in $B$. \PO grabs and proceeds to $W_i$.

\noindent\underline{Case 3 -- Line~$12$:} Let $v\in R\setminus P_0$. 
\PO can always ensure that \PT controls $v$ by never grabbing $v$.
Also in OVPP optional-grabbing, \PO does not need to grab $v$ in order to reach $v$.
Thus if \PO has a strategy such that the token reaches $v$, then he can ensure that the token reaches $v$ and is controlled by \PT. 
Hence, once $v$ is reached, since it is controlled by \PT, she is forced to move from $v$ to a vertex $v'$ in $B$. 
\PO then grabs $v'$ unless he already controls it and moves the token to $W_i$.
\end{proof}

We turn to prove completeness.

\begin{clm}
Suppose that the algorithm terminates at iteration $i$ in Lines~$7$ or~$13$. Then, \PT has a winning strategy from $\zug{v_0, P_0}$.
\end{clm}
\begin{proof}
We first consider the case of terminating in Line~$7$. Thus, every vertex not in $W_i$, including $v_0$, does not have a path to $W_i$. Clearly \PT wins. 

Second, suppose that the algorithm terminates in Line~$13$. We describe a winning \PT strategy from $\zug{v_0, P_0}$ that avoids $W_i$. 
Whenever \PT moves the token, she moves it to an arbitrary vertex not in $W_i \cup B$. When the token reaches $u \in B$, it is following a \PO move, and \PT grabs $u$, and moves to $W_i \cup B$.

We claim that \PT wins when the game starts from $\zug{v_0, P_0}$. Assume towards contradiction that $W_i$ is reached. Since $v_0 \notin W_i$, a visit to $W_i$ must first visit a vertex $u \in B$. We claim that (i) $u$ is in the control of \PT and that (ii) she can move away from $W_i$. Combining (i) and (ii) we reach a contradicting to the assumption that $W_i$ is reached. We conclude the proof by showing that (i) and (ii) hold.
For (ii), note that $u$ has a neighbor not in $W_i \cup B$ otherwise $u$ would have been added to $W_{i+1}$ in Line~$4$ or Line~$12$. 
We turn to prove (i). Suppose first that $u=v_0$. 
If \PO controls $v_0$, then the algorithm terminates in Line~$8$.
Since it does not terminate there, \PT controls $u$. 
Next we consider the case that $u \neq v_0$, thus $u$ has a predecessor $u'$ in the play. 
We distinguish between two cases. 
If $u' \in R$, then $u' \in P_0$ otherwise it would have been added to $W_{i+1}$ in Line~$12$. 
Thus, \PO makes the move from $u'$ to $u$, and \PT grabs $u$ following \PO's move (if she does not control it already).
The second case is when $u' \notin R$. 
Recall that a vertex is in $R$ if all of its neighbors lead to $B$, thus $u' \notin R$ implies that it has at least one neighbor not in $B$.
Note that if $u'$ was in the control of \PT, her strategy would have dictated to proceed to a vertex not in $B$. 
Thus, if \PO makes the move from $u'$ to $u$, and \PT grabs $u$ following \PO's move. 

Finally, note that the algorithm terminates once a fixed point is reached, thus it runs for at most $|V|$ iterations.
\end{proof}
This concludes the proof of the theorem.
\end{proof}
% A pseudocode of the algorithm is given in Appendix~\ref{app:ovpp-optional-grab}.
\begin{algorithm}[t]
\begin{algorithmic}[1]
\State $W_0 = T$, $i=0$
\While {True}
\If{$v_0\in W_i$} \Return Player 1
\EndIf
\State $W_{i+1} = W_i \cup \set{u: N(u) \subseteq W_i}$
\If{$W_i \neq W_{i+1}$} \text{$i:=i+1$; Continue} \EndIf
\State $B := \set{u: N(u) \cap W_i \neq \emptyset}$
\If{$B=\emptyset$} \Return Player 2 \EndIf
\If{$v_0\in B$ and $v_0\in P_0$} \Return Player 1 \EndIf
\State $B' := \set{u \in B: N(u) \subseteq B \cup W_i}$
\If{$B'\neq \emptyset$} $W_{i+1} := W_i \cup B'$; $i:=i+1$; Continue \EndIf
\State $R = \set{u: N(u) \subseteq B}$
\If{$R \setminus P_0 \neq \emptyset$} $W_{i+1} = W_i \cup (R\setminus P_0)$; $i:=i+1$
\Else~ \Return Player 2
\EndIf
\EndWhile
\end{algorithmic}
    \caption{\label{algo:OVPP_optional}Given an OVPP optional-grabbing pawn game $\G = \zug{V, E, T}$ and an initial configuration $c = \zug{v, P_0}$, determines which player wins $\G$ from $c$.}
\end{algorithm}

\subsection{MVPP: EXPTIME-hardness via \LAK games}\label{sec:lnk}
%\paragraph*{\LAK games}
%In this section, we prove that MVPP optional-grabbing pawn games are \EXP-hard. To that end,
We prove hardness of MVPP optional-grabbing pawn games by reduction through a class of games that we introduce and call {\em \LAK} games, and may be of independent interest. 
A \LAK game is $\G = \zug{V, E, T, L, K, \lambda, \kappa}$, where $\zug{V, E, T}$ is a turn-based game, $L = \set{\lock_1,\ldots, \lock_n}$ is a set of locks $K = \set{k_1,\ldots, k_n}$ is a set of keys, each $\lock_j$ is associated to key $\key_j \in K$ for $j \in [n]$, and each edge is labeled by a set of locks and keys respectively given by $\lambda: E \rightarrow 2^L$ and $\kappa: E \rightarrow 2^K$. Note that a lock and a key can appear on multiple edges.

Intuitively, a \LAK game is a turn-based game, only that the locks impose restrictions on the edges that a player is allowed to cross. Formally, a configuration of a \LAK game is $c = \zug{v, A} \in V \times 2^L$, meaning that the token is placed on $v$ and each lock in $A$ is {\em closed} (all other locks are {\em open}). When $v \in V_i$, for $i \in \set{1,2}$, then \PLi moves the token as in turn-based games with the restriction that he cannot choose an edge that is labeled by a closed lock, thus $e = \zug{v, u} \in E$ is a legal move at $c$ when $\lambda(e) \subseteq (L \setminus A)$. Crossing $e$ updates the configuration of the locks by ``turning'' all keys that $e$ is labeled with. Formally, let $\zug{u, A'}$ be the configuration after crossing $e$. For $k_j \in \kappa(e)$ (``key $k_j$ is turned''), we have $\ell_j \in A$ iff $\ell_j \notin A'$. For $k_j \notin \kappa(e)$ (``key $k_j$ is unchanged''), we have $\ell_j \in A$ iff $\ell_j \in A'$.

Note that, similar to pawn games, each \LAK game corresponds to an exponentially sized two-player turn-based game. Thus, membership in $\EXP$ is immediate. For the lower bound, we show a reduction for the problem of deciding whether an {\em alternating polynomial-space Turing machine} (ATM) accepts a given word. 
% See details in Appendix~\ref{app:LAK-EXP}.

\begin{thm}
\label{thm:LAK-EXP}
Given a \LAK game $\G$ and an initial configuration $c$, deciding whether \PO wins from $c$ in $\G$ is $\EXP$-complete.
\end{thm}
\begin{proof}
We briefly describe the syntax and semantics of ATMs, see for example~\cite{Sipser13}, %short \cite[Chapter 10]{Sipser13}, 
for more details. 
An ATM is $\A = \zug{Q, \Gamma, \delta, q_0, q_{acc}, q_{rej}}$, where $Q$ is a collection of states that is partitioned into $Q = Q_1 \cup Q_2$ owned by \PO and \PT respectively, $\Gamma$ is a tape alphabet, $\delta: Q \times \Gamma \rightarrow 2^{Q \times \Gamma \times \set{{\sf L}, {\sf R}}}$ is a transition function, $q_0, q_{acc}, q_{rej} \in Q$ are respectively an initial, accepting, and rejecting states. A configuration of $\A$ is $c = \zug{q, \pos, \zug{\gamma_1,\ldots, \gamma_m}}$, meaning that the control state is $q$, the head position is $\pos$, and $\zug{\gamma_1,\ldots, \gamma_m}$ is the tape content, where $m$ is polynomial in the length of the input word $w$. In order to determine whether $\A$ accepts $w$ we construct a (succinctly-represented) turn-based game over the possible configurations of $\A$, the neighbors of a configuration are determined according to $\delta$, and, for $i \in \set{1,2}$, \PLi moves from states in $Q_i$. We say that $\A$ accepts $w$ iff \PO has a winning strategy from the initial configuration for the target state $q_{acc}$. 

Given $\A$ and $w$, 
% initial tape configuration $\zug{\gamma_1, \ldots, \gamma_m}$, 
we construct a \LAK game $\G = \zug{V, E, T, L, K, \lambda, \kappa}$ and an initial configuration $\zug{v_0, A}$ such that \PO wins from $\zug{v, A}$ in $\G$ iff $w$ is accepted by $\A$. The vertices of $\G$ consist of {\em main} and {\em intermediate} vertices. Consider a configuration $c = \zug{q, \pos, \zug{\gamma_1,\ldots, \gamma_m}}$ of $\A$. We simulate $c$ in $\G$ using $c' = \zug{v, A}$ as follows. First, the main vertices are $Q \times \set{1,\ldots,m} \times \Gamma$ and keep track of the control state and position on the tape. The main vertex that simulates $c = \zug{q, \pos, \zug{\gamma_1,\ldots, \gamma_m}}$ is $v = \zug{q, \pos, \gamma_\pos}$. We define $v \in V_i$ iff $q \in Q_i$. Second, we use locks to keep track of the tape contents. For each $1 \leq i \leq m$ and $\gamma \in \Gamma$, we introduce a lock $\lock_{i, \gamma}$. Then, in the configuration $c' = \zug{v, A}$ that simulates $c$,  %the only control lock that is open is $\lock_q$, the only head position lock that is open is $\lock_p$ and 
the only locks that are open are $\lock_{i, \gamma_i}$, for $i \in \set{1,\ldots,m}$. 
Next, we describe the transitions, where intermediate vertices are used for book-keeping.
The neighbors of a main vertex $v$ are the intermediate vertices $\set{\zug{v, t}: t \in \delta(q, \gamma)}$, where a transition of $\A$ is $t = \zug{q', \gamma', B}$, meaning that the next control state is $q'$, the tape head moves to $\pos+1$ if $B = R$ and to $\pos-1$ if $B=L$, and the $\pos$-th tape content changes from $\gamma$ to $\gamma'$. We update the state of the locks so that they reflect the tape contents: for the edge $\zug{v, \zug{v, t}}$, we have $\kappa(\zug{v, \zug{v, t}}) = \{k_{i, \gamma}, k_{i, \gamma'}\}$. That is, traversing the edge turn the keys to close  $\lock_{\pos, \gamma}$ and open $\lock_{\pos, \gamma'}$.
The neighbors of $\zug{v, t}$ are main vertices having control state $q'$ and head position $\pos'$. Recall that the third component of a main vertex is the tape content at the current position. We use the locks' state to prevent moving to main vertices with incorrect tape content: outgoing edges from $\zug{v,t}$ are of the form $\zug{\zug{v, t}, \zug{q', \pos', \gamma''}}$ and is labeled by the lock $\lock_{\pos', \gamma''}$. That is, the edge can only be traversed when the $\pos'$-th tape position is $\gamma''$. 
It is not hard to verify that there is a one-to-one correspondence between runs of $\A$ and plays of $\G$.
Thus, \PO forces $\A$ to reach a configuration with control state $q_{acc}$ iff \PO forces to reach a main vertex with control state $q_{acc}$.
 Note that the construction is clearly polynomial since $\G$ has $|Q| \cdot m \cdot |\Gamma|$ main vertices. 
\end{proof}

\makeatletter
\def\squarecorner#1{
    \pgf@x=\the\wd\pgfnodeparttextbox%
    \pgfmathsetlength\pgf@xc{\pgfkeysvalueof{/pgf/inner xsep}}%
    \advance\pgf@x by 2\pgf@xc%
    \pgfmathsetlength\pgf@xb{\pgfkeysvalueof{/pgf/minimum width}}%
    \ifdim\pgf@x<\pgf@xb%
        \pgf@x=\pgf@xb%
    \fi%
    \pgf@y=\ht\pgfnodeparttextbox%
    \advance\pgf@y by\dp\pgfnodeparttextbox%
    \pgfmathsetlength\pgf@yc{\pgfkeysvalueof{/pgf/inner ysep}}%
    \advance\pgf@y by 2\pgf@yc%
    \pgfmathsetlength\pgf@yb{\pgfkeysvalueof{/pgf/minimum height}}%
    \ifdim\pgf@y<\pgf@yb%
        \pgf@y=\pgf@yb%
    \fi%
    \ifdim\pgf@x<\pgf@y%
        \pgf@x=\pgf@y%
    \else
        \pgf@y=\pgf@x%
    \fi
    \pgf@x=#1.5\pgf@x%
    \advance\pgf@x by.5\wd\pgfnodeparttextbox%
    \pgfmathsetlength\pgf@xa{\pgfkeysvalueof{/pgf/outer xsep}}%
    \advance\pgf@x by#1\pgf@xa%
    \pgf@y=#1.5\pgf@y%
    \advance\pgf@y by-.5\dp\pgfnodeparttextbox%
    \advance\pgf@y by.5\ht\pgfnodeparttextbox%
    \pgfmathsetlength\pgf@ya{\pgfkeysvalueof{/pgf/outer ysep}}%
    \advance\pgf@y by#1\pgf@ya%
}
\makeatother
\pgfdeclareshape{square}{
    \savedanchor\northeast{\squarecorner{}}
    \savedanchor\southwest{\squarecorner{-}}

    \foreach \x in {east,west} \foreach \y in {north,mid,base,south} {
        \inheritanchor[from=rectangle]{\y\space\x}
    }
    \foreach \x in {east,west,north,mid,base,south,center,text} {
        \inheritanchor[from=rectangle]{\x}
    }
    \inheritanchorborder[from=rectangle]
    \inheritbackgroundpath[from=rectangle]
}

\subsubsection{From \LAK games to optional-grabbing pawn games}
Throughout this section, fix a \LAK game $\G$ and an initial configuration $c$. We construct an optional-grabbing pawn game $\G'$ over a set of pawns $[d]$, and identify an initial configuration $c'$ such that \PO wins in $\G$ from $c$ iff \PO wins from $c'$ in $\G'$.

\paragraph*{From turn-based games to optional-grabbing games}
\begin{figure}[t]
% \begin{wrapfigure}{R}{0.25\textwidth}
\centering
%\scalebox{0.7}{
\begin{tikzpicture}[node distance={8mm},initial text = ,main/.style = {draw, circle, text=black, inner sep=1pt},gain/.style = {draw, square, text=black, inner sep=1pt}] 
%\node[main] (1) [right of=6] {$v'_2$};
%\node[main,initial] (1) {$v_1$};
\node[main,initial] (2) {$v'_1$};
\node[gain,initial] (8) [left of=2, xshift=-2cm] {$v_1$};
\node[main] (9) [right of=8] {$v_2$};
% \node[main,initial] (2) [right of=1] {$v'_1$};
\node[gain] (3) [right of=2] {$v_1$};
\node[gain] (4) [right of=3] {$v'_2$};
\node[main] (5) [right of=4] {$v_2$};
% \node[main,accepting] (6) [below of=3] {$t$};
\node[main] (7) [right of=5] {$s$};
\node[main,accepting] (6) [right of=7] {$t$};
\draw[->] (8) -- (9);
\draw[->] (3) -- (4);
\draw[->] (2) -- (3);
\draw[->] (4) -- (5);
\draw[->] (5) -- (7);
% \draw[->, bend right] (3) -- (6);
 \path[->]
(3) edge[bend right] node [left] {} (6);
% (5) edge[bend right] node [left] {} (7);
\end{tikzpicture}
%}
\caption{From turn-based to optional-grabbing games.}
\label{fig:turnbased-to-OVPP}
% \end{wrapfigure}
\end{figure}
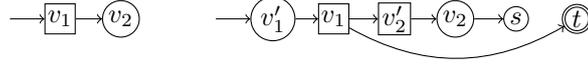
In this section, we consider the case in which $\G$ has no keys or locks, thus $\G$ is a turn-based game.
%Consider a \LAK game $\G$ and an initial configuration $c$. We construct an optional-grabbing pawn game $\G'$ and an initial configuration $c'$ such that \PO wins in $\G$ from $c$ iff he wins in $\G'$ from $c'$. We first consider turn-based games, which are \LAK games with no locks or keys. 
The reduction is depicted in Fig.~\ref{fig:turnbased-to-OVPP}.
Denote the turn-based game $\G = \zug{V, E, T}$ with $V = V_1 \cup V_2$ and initial vertex $v_0$. We construct an OVPP optional-grabbing $\G' = \zug{V', E', T'}$, where $V' = V \cup \set{v': v \in V} \cup \set{s,t}$. We add edges to ensure that the player who owns a vertex $v \in V$ is the player who moves from $v$ in $\G'$: we have $\zug{v', v} \in E'$, and if $v \in V_1$, then $\zug{v, s} \in E'$, and if $v \in V_2$, then $\zug{v, t} \in E'$. 
% For $v \in V$, 
We redirect each edge $\zug{u,v}$ in $\G$ to $\zug{u, v'}$ in $\G'$. Intuitively, for $v \in V_1$, a \PO winning strategy will guarantee that $v'$ is always in the control of \PT, and following her move at $v'$, \PO must grab $v$ otherwise \PT wins and choose the next location. And dually for $v \in V_2$. Let $V'_1 = V_1 \cup \set{v': v \in V_2}$, the initial configuration of $\G'$ is $\zug{v_0, V'_1}$, that is \PT controls $V_2 \cup \set{v': v \in V_1}$. 
% Formally, we prove the following in App.~\ref{app:TB}.
We thus have the following:

%Consider a \LAK game $\G$ and an initial configuration $c$. We construct an optional-grabbing pawn game $\G'$ and an initial configuration $c'$ such that \PO wins in $\G$ from $c$ iff he wins in $\G'$ from $c'$. The first step in the reduction considers turn-based games, which are \LAK games with no locks or keys. Consider a turn-based game $\G = \zug{V, E, T}$ with $V = V_1 \cup V_2$. We construct an OVPP optional-grabbing $\G' = \zug{V', E', T'}$, where $V'$ consists of two copies of $V$ as well as a fresh vertex $s$ that is winning for \PT, i.e., there is no path from $s$ to $T'$. In the initial configuration, \PO controls the vertices $V'_1 = V_1 \cup \set{v': v \in V_2}$ and \PT, the vertices $V'_2 = V_2 \cup \set{v': v \in V_1}$.  We describe $E'$. Let $v \in V$. Incoming edges to $v$ in $\G$ are redirected to $v'$ in $\G'$, thus if $\zug{u, v} \in E$, then $\zug{u, v'} \in E'$. %short Implicitly this defines the outgoing edges from $v$: if $\zug{v, u} \in E$, then $\zug{v, u'} \in E'$. Finally, we add edges to ensure that the player who owns a vertex $v \in V$ is the player who moves from $v$ in $\G'$. Set $\zug{v', v} \in E'$. If $v \in V_1$, then $\zug{v, s} \in E'$, and if $v \in V_2$, then $\zug{v, t} \in E'$. Intuitively, let $v \in V_1$, then when $\G'$ reaches $v'$, it is at the control of \PT, she must move to $v$, which allows \PO to grab $v$. Moreover, \PO must grab $v$ since $s \in N(v)$. Formally, we prove the following in Appendix~\ref{app:TB}.
%Since $\G'$ is an OVPP, by abuse of notation, we refer to a pawn by the vertex it owns.

\begin{lem}
\label{lem:TB}
For a turn-based game $\G$, \PO wins $\G$ from a vertex $v_0 \in V$ iff \PO wins the optional-grabbing game $\G'$ from configuration $\zug{v_0, {V'_1}}$.
\end{lem}
\begin{proof}
Let $f$ be a \PO winning strategy in $\G$. We describe a \PO winning strategy $f'$ in $\G'$. The proof is dual when \PT wins in $\G$. \PO's strategy $f'$ simulates the execution of $f$ on $\G$ so that when the token is placed on $v \in V$ in $\G'$ it is also placed on $v$ in $\G$. Moreover, if $v \in V_1$ in $\G$, then \PO controls $v$ in $\G'$. 

Initially, the invariant clearly holds. Suppose that \PT is playing $\G'$ according to some strategy $g'$. We show that either $f'$ wins by reaching $t$, or that the invariant is maintained. Suppose that the token is placed on $v \in V$. We distinguish between three cases. (1) If $v \in V_2$ and \PO owns $v$ in $\G'$, then he moves to $t$ to win the game. (2) Suppose that $v \in V_2$, that \PT owns $v$, and that she moves to $u'$. 
The simulation in $\G$ proceeds to vertex $u$. 
In order to maintain the second part of the invariant, \PO does not grab $u'$ but grabs $u$ when $u \in V_1$, which is possible because in such a case we define $u' \in V'_2$. (3) When $v \in V_1$, the invariant implies that \PO controls $v$ in $\G'$. \PO then copies the move of $f$ in $\G$; namely, let $u = f(v)$, then $u' = f'(v)$. The move from $u'$ is as in case (2).

Let $v_0, v_1,\ldots$ be the play in $\G$. Clearly, if case (1) occurs at some point, \PO wins $\G'$. Assume that it does not occur. Then, the invariant implies that the play in $\G'$ is $v_0, v'_1, v_1, v'_2, v_2,\ldots$. Since $f$ is winning in $\G$, there is an $i \geq 0$ such that $v_i = t$, thus the play is winning in $\G'$ as well.
\end{proof}

\paragraph*{Gadgets for simulating locks and keys}
The core of the reduction is to construct gadgets that simulate locks and keys. Let $\G'$ denote the optional-grabbing pawn game that we construct, and let $[d]$ denote its set of pawns. 
For each lock $\ell \in L$ and its corresponding key $k \in K$, we construct gadgets $\G_\ell$ and $\G_k$ that simulate the operations of $\ell$ and $k$ in $\G'$. 
The gadgets in two {\em states} are depicted in Fig.~\ref{fig:gadgets}. We highlight three pawns colored blue, green, and red, respectively owning, $\set{v_1^\ell, v_1^k}$, $\set{v_2^\ell, v_2^k, v_7^k, v_8^k}$, and $\set{v_{\text{in}}^k, v_4^k, v_5^k, v_6^k}$. Each of the other vertices (colored white) is owned by a fresh pawn.
Intuitively, for each lock $\ell$, we identify two sets $\P^\ell_O, \P^\ell_C \subseteq 2^{[d]}$, respectively representing an open and closed state of $\ell$.
We will ensure that when entering and exiting a gadget, the configuration is in one of these sets. 
When the set of pawns that \PO controls is in $\P^\ell_O$ and $\P^\ell_C$, we respectively say that $\G_\ell$ is in open and closed state, and similarly for $\G_k$ as stated below. 
We define $\P^\ell_O = \set{P \in 2^{[d]}: v^\ell_1 \notin P \wedge v^\ell_2 \in P}$ and $\P^\ell_C = \set{P \in 2^{[d]}: v^\ell_1 \in P \wedge v^\ell_2 \notin P}$. 
Formally, we define $\P^\ell_O = \set{P \in 2^{[d]}: v^\ell_1 \notin P \wedge v^\ell_2 \in P}$ and $\P^\ell_C = \set{P \in 2^{[d]}: v^\ell_1 \in P \wedge v^\ell_2 \notin P}$.

\begin{lem}
\label{lem:lock}
Let $i \in \set{1,2}$. % and let $P$ and $P'$ be a set of pawns that \PO owns upon entry and exit of $\G_\ell$.
An open lock stays open: If \PLi enters $\G_\ell$ in $\P^\ell_O$, then he has a strategy that guarantees that either he wins $\G'$ or $\G_\ell$ is exited  in $\P^\ell_O$. A closed lock cannot be crossed: If \PLi enters $\G_\ell$ in $\P^\ell_C$, then \PLni has a strategy that guarantees that \PLi loses $\G'$.
\end{lem}
\begin{proof}
%Suppose that \PO enters $\G_\ell$ in a configuration in $\P^\ell_O$. We describe dominant strategies for the two players that guarantee that both players do not lose and that the gadget is exited. We show that the configuration upon exiting $\G_\ell$ is in $\P^\ell_O$. The case when \PT enters $\G_\ell$ is dual. 
We prove for \PO and the proof is dual for \PT. First, suppose \PO enters $\G_\ell$ in $\P^\ell_O$. 
% It is not hard to check that it is dominant for \PT to grab $v^\ell_{\text{in}}$ and proceed to $v^\ell_1$. 
\PT may or may not grab $v^\ell_{\text{in}}$, and the game can proceed to either $v^\ell_1$ or $v^\ell_2$.
We argue that if the game proceeds to $v^\ell_1$, then \PO will not grab $v^\ell_1$.
We can also similarly show that if the game proceeds to $v^\ell_2$, then \PT will not grab $v^\ell_2$.
\PT controls $v^\ell_1$. We claim that if \PO grabs $v^\ell_1$, he will lose the game. Indeed, following \PO's move in $v^\ell_1$, \PT will grab $v^\ell_3$ and move the token to the sink vertex $s$ to win the game.
Thus, \PO does not grab $v^\ell_1$ and keeps it in the control of \PT. Following \PT's move in $v^\ell_1$, \PO grabs $v^\ell_3$ and proceeds to exit $\G_\ell$. 
Note that when $\G_\ell$ is exited, \PO maintains control of $v^\ell_2$ and \PT maintains control of $v^\ell_1$, thus the configuration is in $\P^\ell_O$. Second, suppose that \PO enters $\G_\ell$ in $\P^\ell_C$. Then, \PT grabs $v^\ell_{\text{in}}$ and moves the token to $v^\ell_1$. Since \PO controls $v^\ell_1$ he must make the next move. \PT then grabs $v^\ell_3$ and moves the token to $s$ to win the game.
\end{proof}

Next, we present the gadget $\G_k$ for simulating the operation of a key $k$ (see Fig.~\ref{fig:gadgets}). Intuitively, we maintain that $\G_k$ is in open state iff $\G_\ell$ is in open state, and traversing $\G_k$ swaps the state of both. 
We define sets of configurations $\P^k_O = \set{P \in 2^{[d]}: \set{v^k_{\text{in}}, v^k_1, v^k_4, v^k_5, v^k_6} \cap P = \emptyset \wedge \set{v^k_2, v^k_7, v^k_8} \subseteq P}$ and $\P^k_C = \set{P \in 2^{[d]}: \set{v^k_{\text{in}}, v^k_1, v^k_4, v^k_5, v^k_6} \subseteq P \wedge \set{v^k_2, v^k_7, v^k_8} \cap P = \emptyset}$ (see Fig.~\ref{fig:gadgets}). 
Note that $\P^k_O \subseteq \P^\ell_O$ and $\P^k_C \subseteq \P^\ell_C$ since $v_i^k$ and $v_i^\ell$ are owned by the same pawn for $i \in [2]$.

\begin{lem}
\label{lem:turning-key}
Turning $k$ closes an open $\ell$: Let $i \in \set{1,2}$. If \PLi enters $\G_k$ in $\P^k_O$, then he has a strategy that ensures that either \PLi wins $\G'$ or $\G_k$ is exited in $\P^k_C$. 
Turning $k$ opens a closed $\ell$: when \PLi enters $\G_k$ in $\P^k_C$, \PLi ensures that either he wins $\G'$ or $\G_k$ is exited in $\P^k_O$. 
\end{lem}
\begin{proof}
    We depict $\G_k$ in two configurations in Fig.~\ref{fig:gadgets}. The vertices $v^k_3$ and $v^k_{\text{out}}$ are controlled by one-vertex pawns. The rest of the vertices in $\G_k$, not including targets, are controlled by three pawns and are depicted using colors. Vertex $v^k_1$ is controlled by a ``blue'' pawn, who also owns $v^\ell_1$. Vertices $v^k_2$, $v^k_7$, and $v^k_8$ are all owned by a ``green'' pawn who also controls $v^\ell_2$. The other vertices are owned by a ``red'' pawn. 
    
    We simulate the properties of a key using the configurations, where we prove the claim for an entry configuration in $\P^k_C$ and the proof for $\P^k_O$ is dual. We claim that for $i \in \set{1,2}$, if the token lands on $v^k_2$ when \PLi controls that vertex, then \PLi loses. 
    Indeed, following \PLi's move at $v^k_2$, \PLni grabs $v^k_3$ and directs the token to the winning vertex. 
It follows that \PO does not grab $v^k_{\text{in}}$ upon entry to $\G_k$ and allows \PT to move. Following \PT's move at $v^k_{\text{in}}$, \PO grabs $v^k_1$, and proceeds to $v^k_4$. 
Since \PO moves and $v^k_4$ is in the control of \PT, no change of pawns occurs. We claim that 
% it is dominant move for 
\PT now proceeds to $v^k_5$. Indeed, if she proceeds to $v^k_6$, \PO will grab $v^k_6$ and win by moving to $t$. 
Observe that \PO grabs the red pawn since it controls $v_5^k$ following \PT's move from $v^k_4$ to $v^k_5$.
Indeed, otherwise \PT proceeds to $s$ to win the game.
Finally, following \PO's move from $v^k_5$, \PT must grab the green pawn since it grabs $v_7^k$, otherwise \PO moves from $v^k_7$ to $t$ to win the game.
To conclude, if the token exits $\G_k$, \PO has grabbed the blue and red pawns and \PT has grabbed the green pawn. The configuration upon exiting $\G_k$ is thus in $\P^k_O$, and we are done.
\end{proof}

\usetikzlibrary{automata,arrows}
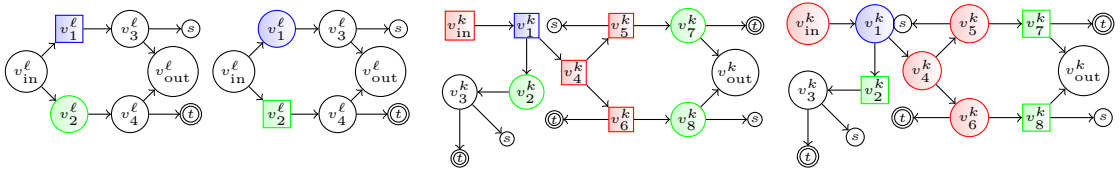
\begin{figure}[t]
\hspace*{-25pt}
 \begin{minipage}[b]{5cm}
\begin{tikzpicture}[node distance={8mm}, main/.style = {draw,circle, text=black, inner sep=1pt},main1/.style = {draw,circle, blue, left color=blue!20, text=black, inner sep=1pt},main2/.style = {draw,circle, red, left color=red!20, text=black, inner sep=1pt},main3/.style = {draw,circle, green, left color=green!20, text=black, inner sep=1pt},gain1/.style = {draw,square, blue, left color=blue!20, text=black, inner sep=1pt},gain2/.style = {draw,square, red, left color=red!20, text=black, inner sep=1pt},gain3/.style = {draw,square, green, left color=green!20, text=black, inner sep=1pt}] ]
\node[main] (1) {\tiny$v^\ell_{\text{in}}$};
\node[gain1] (2) [above right of=1] {\tiny$v^\ell_1$};
\node[main3] (3) [below right of=1] {\tiny$v^\ell_2$};
\node[main] (4) [right of=2] {\tiny$v^\ell_3$};
\node[main] (5) [right of=3] {\tiny$v^\ell_4$};
\node[main] (6) [below right of=4] {\tiny$v^\ell_{\text{out}}$};
\node[main] (7) [right of=4] {\tiny$s$};
\node[main,accepting] (8) [right of=5] {\tiny$t$};
\node[main] (9) [right of=6]{\tiny$v^\ell_{\text{in}}$};
\node[main1] (10) [above right of=9] {\tiny$v^\ell_1$};
\node[main] (12) [right of=10] {\tiny$v^\ell_3$};
\node[main] (13) [below right of=12] {\tiny$v^\ell_{\text{out}}$};
\node[main] (14) [below left of=13] {\tiny$v^\ell_4$};
\node[main] (15) [right of=12] {\tiny$s$};
\node[main,accepting] (16) [right of=14] {\tiny$t$};
\node[gain3] (17) [left of=14] {\tiny$v^\ell_2$};
\draw[->] (1) -- (2);
\draw[->] (1) -- (3);
\draw[->] (2) -- (4);
\draw[->] (3) -- (5);
\draw[->] (4) -- (6);
\draw[->] (5) -- (6);
\draw[->] (4) -- (7);
\draw[->] (5) -- (8);
\draw[->] (9) -- (10);
\draw[->] (9) -- (17);
\draw[->] (10) -- (12);
\draw[->] (12) -- (15);
\draw[->] (12) -- (13);
\draw[->] (14) -- (13);
\draw[->] (14) -- (16);
\draw[->] (17) -- (14);
\end{tikzpicture}\\
\end{minipage}
\hspace{0.4cm}
  \begin{minipage}[b]{4cm}
\begin{tikzpicture}[node distance={8.8mm}, main/.style = {draw,circle, text=black, inner sep=0.5pt},main1/.style = {draw,circle, blue, left color=blue!20, text=black, inner sep=0.5pt},main2/.style = {draw,circle, red, left color=red!20, text=black, inner sep=0.5pt},main3/.style = {draw,circle, green, left color=green!20, text=black, inner sep=0.5pt},gain1/.style = {draw,square, blue, left color=blue!20, text=black, inner sep=0.5pt},gain2/.style = {draw,square, red, left color=red!20, text=black, inner sep=0.5pt},gain3/.style = {draw,square, green, left color=green!20, text=black, inner sep=0.5pt}] ]
\node[gain2] (1) {\tiny$v^k_{\text{in}}$};
\node[gain1] (2) [right of=1] {\tiny$v^k_1$};
\node[main3] (3) [below of=2] {\tiny$v^k_2$};
\node[main] (4) [left of=3] {\tiny$v^k_3$};
\node[main] (5) [below right of=4] {\tiny$s$}; %was {$v_{k_{s3}}$};
\node[accepting,main] (6) [below of=4] {\tiny$t$}; %was {$v_{k_{t3}}$};
%\node[gain4] (20) [right of=2] { };
\node[gain2] (7) [below right of=2] {\tiny$v^k_4$};
\node[gain2] (8) [above right of=7] {\tiny$v^k_5$};
\node[gain2] (9) [below right of=7] {\tiny$v^k_6$};
\node[main3] (10) [right of=8] {\tiny$v^k_7$};
\node[main3] (11) [right of=9] {\tiny$v^k_8$};
\node[main] (12) [below right of=10] {\tiny$v^k_{\text{out}}$};
\node[accepting,main] (13) [right of=10] {\tiny$t$}; %was {$v_{k_{t1}}$};
\node[main] (14) [right of=11] {\tiny$s$}; %was {$v_{k_{s1}}$};
\node[main] (15) [left of=8] {\tiny$s$}; %was {$v_{k_{s2}}$};
\node[accepting,main] (16) [left of=9] {\tiny$t$}; %was {$v_{k_{t2}}$};
\draw[->] (1) -- (2);
\draw[->] (2) -- (3);
\draw[->] (3) -- (4);
\draw[->] (4) -- (5);
\draw[->] (4) -- (6);
\draw[->] (2) -- (7);
\draw[->] (7) -- (8);
\draw[->] (7) -- (9);
\draw[->] (8) -- (10);
\draw[->] (9) -- (11);
\draw[->] (10) -- (12);
\draw[->] (11) -- (12);
\draw[->] (10) -- (13);
\draw[->] (11) -- (14);
\draw[->] (8) -- (15);
\draw[->] (9) -- (16);
\end{tikzpicture}
\end{minipage}
\hspace{0.3cm}
    \begin{minipage}[b]{3.7cm}
\begin{tikzpicture}[node distance={8.8mm}, main/.style = {draw,circle, text=black, inner sep=1pt},main1/.style = {draw,circle, blue, left color=blue!20, text=black, inner sep=1pt},main2/.style = {draw,circle, red, left color=red!20, text=black, inner sep=1pt},main3/.style = {draw,circle, green, left color=green!20, text=black, inner sep=1pt},gain1/.style = {draw,square, blue, left color=blue!20, text=black, inner sep=1pt},gain2/.style = {draw,square, red, left color=red!20, text=black, inner sep=1pt},gain3/.style = {draw,square, green, left color=green!20, text=black, inner sep=1pt},gain4/.style = {text=black}] ]
\node[main2] (1) {\tiny$v^k_{\text{in}}$};
\node[main1] (2) [right of=1] {\tiny$v^k_1$};
\node[gain3] (3) [below of=2] {\tiny$v^k_2$};
\node[main] (4) [left of=3] {\tiny$v^k_3$};
\node[main] (5) [below right of=4] {\tiny$s$}; %was {$v_{k_{s3}}$};
\node[accepting,main] (6) [below of=4] {\tiny$t$}; %was {$v_{k_{t3}}$};
\node[main2] (7) [below right of=2] {\tiny$v^k_4$};
\node[main2] (8) [above right of=7] {\tiny$v^k_5$};
\node[main2] (9) [below right of=7] {\tiny$v^k_6$};
\node[gain3] (10) [right of=8] {\tiny$v^k_7$};
\node[gain3] (11) [right of=9] {\tiny$v^k_8$};
\node[main] (12) [below right of=10] {\tiny$v^k_{\text{out}}$};
\node[accepting,main] (13) [right of=10] {\tiny$t$}; %was {$v_{k_{t1}}$};
\node[main] (14) [right of=11] {\tiny$s$}; %was {$v_{k_{s1}}$};
\node[main] (15) [left of=8] {\tiny$s$}; %was {$v_{k_{s2}}$};
\node[accepting,main] (16) [left of=9] {\tiny$t$}; %was  {$v_{k_{t2}}$};
\draw[->] (1) -- (2);
\draw[->] (2) -- (3);
\draw[->] (3) -- (4);
\draw[->] (4) -- (5);
\draw[->] (4) -- (6);
\draw[->] (2) -- (7);
\draw[->] (7) -- (8);
\draw[->] (7) -- (9);
\draw[->] (8) -- (10);
\draw[->] (9) -- (11);
\draw[->] (10) -- (12);
\draw[->] (11) -- (12);
\draw[->] (10) -- (13);
\draw[->] (11) -- (14);
\draw[->] (8) -- (15);
\draw[->] (9) -- (16);
\end{tikzpicture}
\end{minipage}
\caption{From left to right: $\G_\ell$ in open and closed state and $\G_k$ in open and closed state.}
\label{fig:gadgets}
\end{figure}

\paragraph*{Putting it all together}
We describe the construction of a pawn game $\G'$ from a \LAK game $\G$. We assume w.l.o.g. that each edge $\zug{u, v}$ in $\G$ is labeled by at most one lock or key since an edge that is labeled by multiple locks or keys can be split into a chain of edges, each labeled by a single lock or a key. 
We describe the construction of $\G'$.  
We first apply the construction for turn-based games on $\G$ while ``ignoring'' the locks and keys. Recall that the construction introduces fresh vertices so that an edge $e = \zug{u,v}$ in $\G$ is mapped to an edge $e' = \zug{u', v}$ in $\G'$. We re-introduce the locks and keys so that the labeling of $e'$ coincides with the labeling of $e$. Next, we replace an edge $e'$ that is labeled by a lock $\ell$, by a copy of $\G_\ell$, and if $e$ is labeled by a key $k$, we replace $e'$ by a copy of $\G_k$. Note that multiple edges could be labeled by the same lock $\ell$. In such a case we use fresh vertices in each copy of $\G_\ell$, but crucially, all gadgets share the same pawns so that they share the same state. And similarly for keys. For an illustration of this construction, see Fig.~\ref{fig:delta-path}, which applies the construction on a \LAK game that is output from the reduction in Thm.~\ref{thm:LAK-EXP}.

Finally, given an initial configuration $c = \zug{v, A}$ of $\G$ we define an initial configuration $c' = \zug{v, P}$ of $\G'$. Note that the initial vertex is the entry point of the gadget that simulates $v$ in $\G'$. For each lock $\ell$ and corresponding key $k$, if $\ell$ is open according to $A$, then $P \in \P^\ell_O$, i.e., both $\G_\ell$ and $\G_k$ are initially in open state. And similarly when $\ell$ is closed according to $A$. 
Combining the properties in Lemmas~\ref{lem:TB},~\ref{lem:lock}, and~\ref{lem:turning-key} implies that \PO wins $\G$ from $c$ iff \PO wins $\G'$ from   $c'$. Thus, by Thm.~\ref{thm:LAK-EXP}, we have the following.

\begin{thm}
\label{thm-MVPP optional grab}
MVPP optional-grabbing PAWN-GAMES is EXPTIME-complete. 
\end{thm}

% \newpage
\section{Always-Grabbing Pawn Games} \label{sec:always_grab}
In this section, we study always-grabbing pawn games and show that MVPP always-grabbing pawn-games are $\EXP$-complete. The main challenge is proving the lower bound. We proceed as follows. 
%short Using Theorem~\ref{thm:exp} and the fact that OMVPP is a generalization of MVPP, it follows that OMVPP always-grabbing pawn-games too are $\EXP$-complete. The lower bound is obtained as follows. 
Let $\M$ be an ATM. Apply the reduction in Thm~\ref{thm:LAK-EXP} and the one in Thm.~\ref{thm-MVPP optional grab} to obtain pairs $\zug{\G, c}$ and $\zug{\G', c'}$, where $\G$ and $\G'$ are respectively \LAK and optional-grabbing games with initial configurations $c$ and $c'$ respectively.
We devise a construction that takes $\zug{\G', c'}$ and produces an always-grabbing game $\G''$ and a configuration $c''$ such that \PO wins from $c''$ in $\G''$ iff he wins from $c$ in $\G$.
%We describe the idea behind our construction of $\G''$ from $\G'$.
%Let us first recall the game arena of $\G'$.

\usetikzlibrary{shapes.geometric}
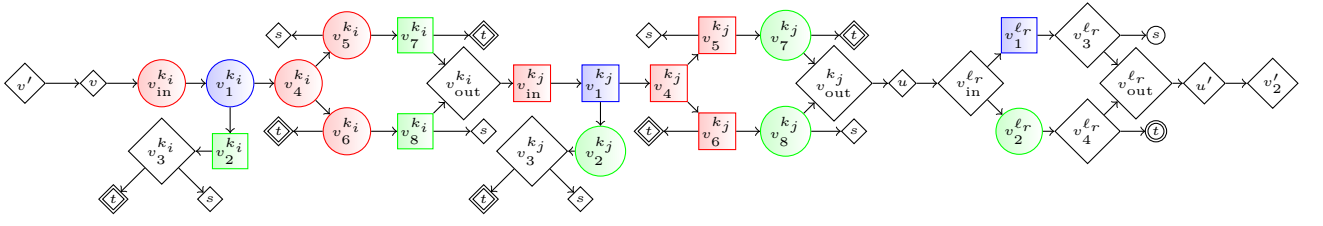
\begin{figure}[t]
\begin{tikzpicture}[node distance={9mm}, main/.style = {draw,circle, text=black, inner sep=1pt},gain/.style = {draw,square, text=black, inner sep=1pt},main1/.style = {draw,circle, blue, left color=blue!20, text=black, inner sep=1pt},main2/.style = {draw,circle, red, left color=red!20, text=black, inner sep=1pt},main3/.style = {draw,circle, green, left color=green!20, text=black, inner sep=1pt},gain1/.style = {draw,square, blue, left color=blue!20, text=black, inner sep=1pt},gain2/.style = {draw,square, red, left color=red!20, text=black, inner sep=1pt},gain3/.style = {draw,square, green, left color=green!20, text=black, inner sep=1pt},gain/.style = {draw,square, text=black, inner sep=1pt},pain/.style = {draw,diamond, text=black, inner sep=1pt}] 
\node[main2] (1) {\tiny$v^{k_i}_{\text{in}}$};
\node[main1] (2) [right of=1] {\tiny$v^{k_i}_1$};
\node[gain3] (3) [below of=2] {\tiny$v^{k_i}_2$};
\node[pain] (4) [left of=3] {\tiny$v^{k_i}_3$};
\node[pain] (5) [below right of=4] {\tiny$s$}; %was {$v_{k_{s3}}$};
\node[accepting,pain] (6) [below left of=4] {\tiny$t$}; %was {$v_{k_{t3}}$};
\node[main2] (7) [right of=2] {\tiny$v^{k_i}_4$};
\node[main2] (8) [above right of=7] {\tiny$v^{k_i}_5$};
\node[main2] (9) [below right of=7] {\tiny$v^{k_i}_6$};
\node[gain3] (10) [right of=8] {\tiny$v^{k_i}_7$};
\node[gain3] (11) [right of=9] {\tiny$v^{k_i}_8$};
\node[pain] (12) [below right of=10] {\tiny$v^{k_i}_{\text{out}}$};
\node[accepting,pain] (13) [right of=10] {\tiny$t$}; %was {$v_{k_{t1}}$};
\node[pain] (14) [right of=11] {\tiny$s$}; %was {$v_{k_{s1}}$};
\node[pain] (15) [left of=8] {\tiny$s$}; %was {$v_{k_{s2}}$};
\node[accepting,pain] (16) [left of=9] {\tiny$t$}; %was {$v_{k_{t2}}$};
\node[gain2] (111) [right of=12]{\tiny$v^{k_j}_{\text{in}}$};
\node[gain1] (112) [right of=111] {\tiny$v^{k_j}_1$};
\node[main3] (113) [below of=112] {\tiny$v^{k_j}_2$};
\node[pain] (114) [left of=113] {\tiny$v^{k_j}_3$};
\node[pain] (115) [below right of=114] {\tiny$s$}; %was {$v_{k_{s3}}$};
\node[accepting,pain] (116) [below left of=114] {\tiny$t$}; %was {$v_{k_{t3}}$};
\node[gain2] (117) [right of=112] {\tiny$v^{k_j}_4$};
\node[gain2] (118) [above right of=117] {\tiny$v^{k_j}_5$};
\node[gain2] (119) [below right of=117] {\tiny$v^{k_j}_6$};
\node[main3] (1110) [right of=118] {\tiny$v^{k_j}_7$};
\node[main3] (1111) [right of=119] {\tiny$v^{k_j}_8$};
\node[pain] (1112) [below right of=1110] {\tiny$v^{k_j}_{\text{out}}$};
\node[accepting,pain] (1113) [right of=1110] {\tiny$t$}; %was {$v_{k_{t1}}$};
\node[pain] (1114) [right of=1111] {\tiny$s$}; %was {$v_{k_{s1}}$};
\node[pain] (1115) [left of=118] {\tiny$s$}; %was {$v_{k_{s2}}$};
\node[accepting,pain] (1116) [left of=119] {\tiny$t$}; %was  {$v_{k_{t2}}$};
\node[pain] (20) [right of=1112] {\tiny$u$};
\node[pain] (31) [left of=1] {\tiny$v$};
\node[pain] (32) [above of=31] {\tiny$v'$};
% I changed the layout a bit, so the diagram fits on the page. If you do not agree with this, let us know
\node[pain] (21) [below of=20]{\tiny$v^{\ell_r}_{\text{in}}$};
\node[gain1] (22) [above right of=21] {\tiny$v^{\ell_r}_1$};
\node[main3] (23) [below right of=21] {\tiny$v^{\ell_r}_2$};
\node[pain] (24) [right of=22] {\tiny$v^{\ell_r}_3$};
\node[pain] (25) [right of=23] {\tiny$v^{\ell_r}_4$};
\node[pain] (26) [below right of=24] {\tiny$v^{\ell_r}_{\text{out}}$};
\node[main] (27) [right of=24] {\tiny$s$};
\node[main,accepting] (28) [right of=25] {\tiny$t$};
\node[pain] (30) [right of=26] {\tiny$u'$};
\node[pain] (33) [right of=30] {\tiny$v_2'$};
\draw[->] (30) -- (33);
\draw[->] (20) -- (21);
\draw[->] (26) -- (30);
\draw[->] (21) -- (22);
\draw[->] (21) -- (23);
\draw[->] (22) -- (24);
\draw[->] (23) -- (25);
\draw[->] (24) -- (26);
\draw[->] (25) -- (26);
\draw[->] (24) -- (27);
\draw[->] (25) -- (28);
\draw[->] (1112) -- (20);
\draw[->] (12) -- (111);
\draw[->] (31) -- (1);
\draw[->] (32) -- (31);
\draw[->] (111) -- (112);
\draw[->] (112) -- (113);
\draw[->] (113) -- (114);
\draw[->] (114) -- (115);
\draw[->] (114) -- (116);
\draw[->] (112) -- (117);
\draw[->] (117) -- (118);
\draw[->] (117) -- (119);
\draw[->] (118) -- (1110);
\draw[->] (119) -- (1111);
\draw[->] (1110) -- (1112);
\draw[->] (1111) -- (1112);
\draw[->] (1110) -- (1113);
\draw[->] (1111) -- (1114);
\draw[->] (118) -- (1115);
\draw[->] (119) -- (1116);
\draw[->] (1) -- (2);
\draw[->] (2) -- (3);
\draw[->] (3) -- (4);
\draw[->] (4) -- (5);
\draw[->] (4) -- (6);
\draw[->] (2) -- (7);
\draw[->] (7) -- (8);
\draw[->] (7) -- (9);
\draw[->] (8) -- (10);
\draw[->] (9) -- (11);
\draw[->] (10) -- (12);
\draw[->] (11) -- (12);
\draw[->] (10) -- (13);
\draw[->] (11) -- (14);
\draw[->] (8) -- (15);
\draw[->] (9) -- (16);
\end{tikzpicture}\\
\caption{A $\delta$-path is a path between two primed main vertices in an optional- or always-grabbing game, and it crosses two key gadgets and one lock gadget.}
\label{fig:delta-path}
\end{figure}

Our analysis heavily depends on the special structure of $\G'$. The construction in Thm.~\ref{thm:LAK-EXP} outputs a game $\G$ with main vertices of the form $\zug{q,i,\gamma}$ ($q$ is a state, $i$ is a tape position, and $\gamma$ is a letter in the tape alphabet). A play of $\G$ can be partitioned into paths between main vertices. 
Each such path corresponds to one transition of the Turing machine and traverses two keys and a lock before again reaching a main vertex.
Recall that when constructing $\G'$ from $\G$, we replace locks and keys with their respective gadgets, and for every vertex $v$ that belongs to $\G$, we add a new primed vertex $v'$ such that if $v$ is controlled by \PLi then $v'$ is controlled by \PLni. 
We call a path in $\G'$ that corresponds to a path in $\G$ between two successive main vertices, say $v$ and $v'$, a {\em $\delta$-path}. Fig.~\ref{fig:delta-path} 
% in Appendix~\ref{app:delta path} 
depicts a $\delta$-path.
The $\delta$-path shown here is along a closed key $k_i$, an open key $k_j$ and an open lock $\ell_r$ such that $r\neq i$.
An open key represents that the corresponding lock is currently open, and going through this key closes the lock and also the state of the key changes from open to closed.
Similarly, a closed key represents that the corresponding lock is currently closed, and going through this key opens the lock and also the state of the key changes from closed to open.
Recall from the proof of Thm~\ref{thm:LAK-EXP}, that we go through the open key for lock $\ell_{i, \gamma'}$ and the closed key for lock $\ell_{i, \gamma}$ and finally, the open lock $\ell_{i', \gamma''}$.
For simplicity of notation, here, we refer to the keys corresponding to the locks $\ell_{i, \gamma'}$ and $\ell_{i, \gamma}$ as $k_i$ and $k_j$ respectively, while we denote by $\ell_r$ the lock $\ell_{i', \gamma''}$.
Recall from Section \ref{sec:lnk} that the gadget for Key $k_m$ and Lock $\ell_m$ mainly have vertices owned by three pawns $p_{Red_m},p_{Blue_m},p_{Green_m}$.
$p_{Red_m}$ owns vertices $\{v_{in}^{k_m},v_{4}^{k_m},v_{5}^{k_m},v_{6}^{k_m}\}$, pawn  $p_{Blue_m}$ owns $\{v_{1}^{k_m},v_{1}^{l_m}\}$, and pawn $p_{Green_m}$ owns $\{v_{2}^{k_m},v_{7}^{k_m},v_{8}^{k_m},v_{2}^{l_m}\}$.
In this $\delta$-path we have pawns $p_{Red_m},p_{Blue_m},p_{Green_m}$ for $m=i,j,r$ (The pawns $p_{Blue_i},p_{Blue_j},p_{Blue_r}$ are different pawns even though the vertices owned by them have the same colour in a $\delta$-path and the same holds for vertices belonging to pawns $p_{Red}$ and $p_{Green}$.).
The ownership of the diamond vertices is not fixed; they can be either controlled by \PO or \PT.

An important property of the specific optional-grabbing game $\G'$ that is constructed in Thm~\ref{thm:LAK-EXP} from an ATM is that every play of $\G'$ consists of a sequence of $\delta$-paths.
% More details on $\delta$-paths can be found in Section~\ref{app:delta path}.
The following observation can easily be verified:

\begin{obs}
A $\delta$-path from $v'$ to $v_2'$ consists of 20 turns.
\end{obs}

% Now consider the following lemma about a $\delta$-path in an optional-grabbing game.
The following lemma is crucial for the construction of %short the always-grabbing game 
$\G''$. 
% See App.~\ref{app:20_10_rounds} for its proof.
The proof of this lemma is obtained by proving several claims made below.
\begin{lem} \label{lem:20_10_rounds}
For $i \in \set{1,2}$, if \PLi has a strategy in the optional-grabbing game $\G'$ to cross a $\delta$-path from $v'$ to $v_2'$, then \PLi has a strategy that moves the token in at least $10$ rounds and \PLni moves the token in at most 10 rounds in the $\delta$-path.
% Thus, applying the same strategy to the always-grabbing game, whenever \PLi does not grab in the optional-grabbing game, \PLni is in possession of a pawn that \PLi can grab in the always-grabbing game.
\end{lem}
\begin{proof}
We will prove this lemma for $i=1$ that is for \PO.
The proof for \PT is similar.
This follows from the fact that in Figure~\ref{fig:delta-path}, the vertices controlled initially by \PO in Key $k_i$ have their corresponding vertices in Key $k_j$ that are initially controlled by \PT.
Further, the gadget for the Lock $\ell_r$ is symmetric for both players.

In order to prove the lemma we first prove some claims.
Note that in all these claims, we talk about a particular player controlling a vertex, for example, in Claim~\ref{lemma-ag1}, we say that \PT controls $v_{1}^{l_r}$, because \PO can ensure that the game reaches such configuration from vertex $v'$. 
%Otherwise, if \PO controls this vertex, then he loses immediately.
We can indeed show that \PO has a strategy such that in the $\delta$-path, \PT controls vertex $v_{1}^{l_r}$ when the token reaches this vertex.
Refer to Figure~\ref{fig:delta-path} in the following claims.
\begin{clm}\label{lemma-ag1}
If \PT controls vertex $v_{1}^{l_r}$ then \PO has a strategy from $v_{1}^{l_r}$ to reach $v_2'$ which ensures that out of the 4 rounds that moves the token to $v_2'$, he moves the token in at least 2 rounds.
\end{clm}
\begin{proof}
Suppose the token is at vertex $v_{1}^{l_r}$ and \PT controls $v_{1}^{l_r}$.
Then \PT moves the token from $v_{1}^{l_r}$ to $v_{3}^{l_r}$.
\PO can now grab a pawn.
If \PT controls vertex  $v_{3}^{l_r}$, \PO grabs the pawn owning this vertex else he chooses to not grab a pawn.
Note that under both the cases by the end of this round \PO controls vertex $v_{3}^{l_r}$. He now moves the token to $v_{out}^{l_r}$.
Note that till now \PO has moved the token in one round.
Now by the end of this round there are two possible cases:
\begin{enumerate}
    \item { \PO controls $v_{out}^{l_r}$.} In this case \PO moves the token to $u'$, and thus it \PO has moved the token in two rounds, and hence regardless of what happens in next round the claim holds.
    \item { \PT controls $v_{out}^{l_r}$.} In this case \PT moves the token to $u'$ and now \PO can control $u'$ and move the token to $v_2'$, which will be his second instance of moving the token and thus the claim holds.
\end{enumerate}
Hence showed.
\end{proof}
\begin{clm}\label{lemma-ag2}
If \PO controls vertex $v_{2}^{l_r}$ then he has a strategy from $v_{2}^{l_r}$ to reach $v_2'$ which ensures that, out of the 4 rounds that moves the token from $v_2^{l_r}$ to $v_2'$, he moves the token in at least 2 rounds.
\end{clm}
\begin{proof}
Suppose the token is at vertex $v_{2}^{l_r}$ and \PO controls $v_{2}^{l_r}$.
Then \PO moves the token from $v_{2}^{l_r}$ to $v_{4}^{l_r}$.
Now if \PO has the pawn owning vertex $v_{4}^{l_r}$, then \PT is forced to grab this pawn in this round. Thus by the end of this round \PT controls $v_{4}^{l_r}$. 
Now \PT moves the token from $v_{4}^{l_r}$ to $v_{out}^{l_r}$. 
\PO now can grab a pawn. If \PT controls vertex  $v_{out}^{l_r}$, \PO grabs the pawn owning this vertex else he chooses to not grab a pawn.
Note under both the cases by the end of this round \PO controls vertex $v_{out}^{l_r}$. He now moves the token to $u'$, which will be his second instance of moving the token and thus the claim holds.
\end{proof}
\begin{clm}\label{lemma-ag3}
If \PO controls vertex $v_{out}^{k_j}$ then he has a strategy from $v_{out}^{k_j}$ to reach $v_2'$ which ensures that, out of the 7 rounds that moves the token from $v_{out}^{k_j}$ to $v_2'$ he moves the token in at least 4 rounds.
\end{clm}
\begin{proof}
Suppose the token is at vertex $v_{out}^{k_j}$ and \PO controls $v_{out}^{k_j}$.
Then \PO moves the token from $v_{out}^{k_j}$ to $u$.
Now by the end of this round there are two possible cases:
\begin{enumerate}
    \item { \PO controls $u$.} In this case \PO moves the token to $v_{in}^{l_r}$. 
    Note that till now \PO has moved the token in two rounds. Now again by the end of this round there are two possible cases:
    \begin{enumerate}
        \item { \PO controls $v_{in}^{l_r}$.} In this case \PO moves the token to $v_1^{l_r}$. Now note that the token is at vertex $v_1^{l_r}$ and \PT controls $v_1^{l_r}$. Thus by Claim \ref{lemma-ag1} we know that from here \PO moves the token in at least 2 rounds before reaching $v_2'$.
        As till now when the token is at $v_1^{l_r}$ \PO has moved the token in three rounds, thus overall under this case \PO moves the token in at least 5 rounds.
        \item { \PT controls $v_{in}^{l_r}$.} In this case \PT either moves the token to the vertex $v_{1}^{l_r}$ in which case \PO chooses to not grab. 
        Now note that the token is at vertex $v_1^{l_r}$ and \PT controls $v_1^{l_r}$. Thus by Claim \ref{lemma-ag1} we know that from here \PO moves the token in at least 2 rounds before reaching $v_2'$.
        Otherwise suppose from $v_{in}^{l_r}$ \PT moves the token to vertex $v_{2}^{l_r}$, in which case \PO chooses to not grab. 
        Now note that here the token is at vertex $v_2^{l_r}$ and \PO controls $v_2^{l_r}$. Thus by Claim \ref{lemma-ag2} we know that from here \PO moves the token in at least 2 rounds before reaching $v_2'$.
        Thus in this case from $v_{in}^{l_r}$ \PO moves the token in at least two rounds. 
        As till reaching $v_{in}^{l_r}$ \PO has moved the token in two rounds, thus overall under this case \PO moves the token in at least 4 rounds as the token reaches $v_2'$.
    \end{enumerate}
    \item { \PT controls $u$.} In this case \PT moves the token to $v_{in}^{l_r}$.
    \PO now can grab a pawn. If \PT controls the vertex  $v_{in}^{l_r}$, \PO grabs the pawn owning this vertex else he choose to not grab a pawn.
    Note under both the cases by the end of this round \PO controls vertex $v_{in}^{l_r}$. 
    He now moves the token to $v_1^{l_r}$. Now note that the token is at vertex $v_1^{l_r}$ and \PT controls $v_1^{l_r}$ thus by Claim \ref{lemma-ag1} we know that from here \PO moves the token in at least two rounds before reaching $v_2'$. 
    As till reaching $v_1^{l_r}$ \PO has moved the token in two rounds, thus overall under this case \PO moves the token in at least 4 rounds.
\end{enumerate}
Hence showed.
\end{proof}

\begin{clm}\label{lemma-ag4}
If \PT controls vertex $v_{in}^{k_j}$, then \PO has a strategy from $v_{in}^{k_j}$ to reach $v_2'$ which ensures that, out of the 12 rounds that moves the token from $v_{in}^{k_j}$ to $v_2'$ he moves the token in at least 6 rounds.
\end{clm}
\begin{proof}
Suppose the token is at vertex $v_{in}^{k_j}$ and \PT controls $v_{in}^{k_j}$.
Then \PT moves the token from $v_{in}^{k_j}$ to $v_1^{k_j}$. In this round \PO grabs pawn $p_{Blue_j}$ inorder to control $v_1^{k_j}$.
Now \PO controls $v_1^{k_j}$, he moves the token to vertex $v^{k_j}_4$.
In the next round \PT moves the token from $v^{k_j}_4$ to $v^{k_j}_5$. \PO in this round grabs $p_{Red_j}$ inorder to control $v^{k_j}_5$. Now \PO moves the token from $v^{k_j}_5$ to $v^{k_j}_7$. Here \PT grabs $p_{Green_j}$ inorder to control $v^{k_j}_7$. Now \PT moves the token from $v^{k_j}_7$ to $v^{k_j}_{out}$. \PO now can grab a pawn. If \PT controls vertex  $v^{k_j}_{out}$, \PO grabs the pawn owning this vertex else he choose to not grab a pawn.
Note under both the cases by the end of this round \PO controls vertex $v_{out}^{k_j}$. 
Now note that the token is at vertex $v_{out}^{k_j}$ and \PO controls $v_{out}^{k_j}$. Thus by Claim \ref{lemma-ag3} we know that from here \PO moves the token in at least 4 rounds before reaching $v_2'$.
As till reaching $v_{out}^{k_j}$ \PO has moved the token in two rounds, thus overall under this case \PO moves the token in at least 6 rounds.
\end{proof}
\begin{clm}\label{lemma-ag5}
If \PO controls vertex $v_{in}^{k_i}$, then he has a strategy from $v_{in}^{k_i}$ to reach $v_2'$ which ensures that, out of the 18 rounds that moves the token to $v_2'$ he moves the token in at least 9 rounds.
\end{clm}
\begin{proof}
\PO controls vertex $v_{in}^{k_i}$, he moves the token to $v_1^{k_i}$, now \PT grabs $p_{Blue_i}$ in order to control $v_1^{k_i}$.
In the next round \PT moves the token to $v_4^{k_i}$. In this round \PO does not grab a pawn.
Now since $p_{Red_i}$ belongs to \PO and $p_{Red_i}$ owns $v_4^{k_i}$, \PO moves the token to $v_6^{k_i}$.
In this round \PT is forced to grab $p_{Red_i}$ in order to control $v_6^{k_i}$.
Now since here \PT controls $v_6^{k_i}$, she moves the token to $v_8^{k_i}$. Now \PO grabs $p_{Green_i}$ and controls $v_8^{k_i}$.
In the next round \PO moves the token to $v_{out}^{k_i}$.
Now observe that till now \PO has moved the token in three rounds.
Now by the end of this round there two possible cases:
\begin{enumerate}
    \item { \PT controls $v_{out}^{k_i}$.} In this case \PT moves the token to $v_{in}^{k_j}$, \PO here chooses to not grab a pawn. 
    Now observe that the token is at vertex $v_{in}^{k_j}$ and \PT controls vertex $v_{in}^{k_j}$ thus by claim \ref{lemma-ag4} we know that from here \PO moves the token in at least 6 rounds before reaching $v_2'$.
    As until reaching $v_{in}^{k_j}$ \PO has moved the token in three rounds, thus overall under this case \PO moves the token in at least 9 rounds.
    \item { \PO controls $v_{out}^{k_i}$.} In this case \PO moves the token to $v_{in}^{k_j}$. Now note that \PT controls $v_{in}^{k_j}$. 
    Thus the token is at the vertex $v_{in}^{k_j}$ and \PT controls vertex $v_{in}^{k_j}$.
    Now by Claim \ref{lemma-ag4}, we know that from here \PO moves the token in at least 6 rounds before reaching $v_2'$. 
    As until reaching $v_{in}^{k_j}$ \PO has moved the token in four rounds, thus overall under this case \PO moves the token in at least 10 rounds.
\end{enumerate}
Hence showed.
\end{proof}
Now let us prove the lemma.
So the token is at vertex $v'$.
There are two possible cases:
\begin{enumerate}
    \item { \PO controls $v'$.} In this case \PO moves the token to  $v$. Now regardless of who controls, $v$ the token reaches $v_{in}^{k_i}$ with \PO controlling vertex $v_{in}^{k_i}$. Note that \PT will not control $v_{in}^{k_i}$ as he would like to control $v_{1}^{k_i}$. As till reaching $v_{in}^{k_i}$ \PO moved the token in atleast one round and by Claim \ref{lemma-ag5} we know that from here \PO moves the token in at least 9 rounds before reaching $v_2'$, thus overall \PO moves in at least 10 rounds before reaching $v_2'$.
    \item { \PT control $v'$.} In this case \PT moves the token to $v$.
    \PO now can grab a pawn. 
    If \PT controls vertex  $v$, \PO grabs the pawn owning vertex $v$ else he chooses to not grab a pawn.
    Note under both the cases by the end of this round \PO controls the vertex $v'$. He now moves the token to $v_{in}^{k_i}$. Again note that \PT will not control $v_{in}^{k_i}$ as he would like to control $v_{1}^{k_i}$. 
    As until reaching $v_{in}^{k_i}$ \PO moved the token in one round and by Claim \ref{lemma-ag5} we know that from here \PO moves the token in at least 9 rounds before reaching $v_2'$. Thus overall \PO moves in at least 10 rounds before reaching $v_2'$.
\end{enumerate}
% Hence proved.
This concludes the proof of the lemma.
\end{proof}

Let $\G' = \zug{V', E', T'}$ with $d$ pawns. The game $\G''$ is constructed from $\G'$ by adding $2(d+10)$ fresh isolated vertices each owned by a fresh unique pawn. Formally, $\G'' = \zug{V'', E', T'}$, where $V'' = V' \cup \{v_1,v_2,\dots,v_{2(d+10)}\}$ such that $v_j \notin V'$, for $j \in [2(d+10)]$. Consider a configuration $c' = \zug{v, P}$ in $\G'$. Let $c'' = \zug{v, P \cup \set{1,2,\ldots,d+10}}$ be a configuration in $\G''$. 
Note that Lemma~\ref{lem:20_10_rounds} also applies to the always-grabbing game $\G''$, and we get the following.

\begin{cor} \label{cor:grab}
For $i \in \set{1,2}$, if \PLi has a strategy in the always-grabbing game $\G''$ to cross a $\delta$-path from $v'$ to $v_2'$, then \PLi has a strategy such that out of the $20$ rounds in the $\delta$-path, the following hold.
% \begin{inparaenum} 
\begin{enumerate}
% \item \PLni \emph{moves} the token in at most $10$ rounds.
\item \PLni \emph{grabs} a pawn in at least $10$ rounds, and
\item \PLi \emph{grabs} a pawn in at most $10$ rounds.
\end{enumerate}
% \end{inparaenum}
\end{cor}
% We explain the first item in.~\ref{cor:grab}.

Corollary~\ref{cor:grab} follows directly from Lemma~\ref{lem:20_10_rounds} since in an always-grabbing game, the number of times \PLni grabs equals the number of times \PLi moves.
% along a $\delta$-path.
% If \PLi has a strategy to reach $v_2'$ from $v'$ in the $\delta$-path, then by Lem.~\ref{lem:20_10_rounds}, he has one in which he moves in at least $10$ rounds.
% Thus \PLni moves in at most $10$ rounds.
In the remaining part of this section, we show that \PO wins $\G'$ from $c'$ iff \PO wins $\G''$ from the configuration $c''$ described above.

\begin{lem} \label{lem:always-grab_optional-grab}
For $i \in \set{1,2}$, \PLi wins from $c'$ in the optional-grabbing game $\G'$ iff he wins from $c''$ in the always-grabbing game $\G''$.
\end{lem}
\begin{proof}
We first give an outline of the proof before proceeding to proving the lemma formally.
We prove that if \PO has a winning strategy $f'$ in $\G'$ from $c'$, then he has a winning strategy $f''$ from $c''$ in $\G''$.
The case for \PT is analogous and the other direction follows from determinacy~(Thm.~\ref{thm:turn-based}.
% , Appendix~\ref{app:always-grab-equiv}). 
We construct $f''$ to mimic $f'$ with the following difference. 
Whenever $f'$ chooses not to grab, in order to follow the rules of the always-grabbing mechanism, $f''$ grabs a pawn owning an isolated vertex. 
This is possible since we show that we maintain the invariant that along a play in $\G''$ that consists of sequences of $\delta$-paths, at the beginning of each $\delta$-path, \PT has at least $10$ isolated pawns. 
Note that the invariant holds initially due to the definition of $c''$. 
We show that it is maintained. 
Recall from the proof of Theorem~\ref{thm:LAK-EXP} that crossing a $\delta$-path simulates a transition in the Turing machine.
Since $\PO$ has a winning strategy in $\G'$, in a winning play, the strategy enables her to cross the $\delta$-path.
Thus, by Lem.~\ref{lem:20_10_rounds}, \PO moves in at least $10$ rounds. 
Thus, \PT moves in at most $10$ rounds, and during each such round, \PO grabs a pawn. 
Hence, \PO grabs at most $10$ times which thus maintains the invariant.
% In Appendix~\ref{app:always-grab-equiv}, 
We show that $f''$ is a winning \PO strategy.

We now formally prove each of the claims stated above.
Now since for \PO, grabbing an isolated pawn serves no extra purpose than grabbing nothing, the action of not grabbing in the optional-grabbing game can be replaced with grabbing a pawn owning an isolated vertex in the always-grabbing game $\G''$.
% Now if we show that in every round in $G''$ where \PT moves the token, \PT has an isolated pawn for \PO to grab, then this is same as saying that in every round in $G''$ where \PT moves the token, \PO has a choice to not grab anything.
Now assuming that \PO has a winning strategy $f'$ in the optional-grabbing game $\G'$, consider a winning play $\pi'$ for \PO in $\G'$.
Now assume that in every round in which \PO does not grab a pawn in $\G'$ can be replaced with \PO grabbing a pawn owning an isolated vertex in the always-grabbing game $\G''$.
Thus consider a strategy $f''$ of \PO and a strategy of \PT such that in the resulting play \PO grabs a pawn owning an isolated vertex from \PT if in the corresponding round in the optional-grabbing game he does not grab anything, otherwise, $f''$ follows the moves of $f'$.
% We call the play in $G''$ to be $\pi''$.
Now consider that \PO chooses strategy $f''$.
Recall that $f'$ is winning for \PO in $\G'$.
We argue that \PT cannot win in $\G''$ against the strategy $f''$ of \PO.
Suppose \PT uses a strategy and grabs some pawns owning non-isolated vertices while playing against strategy $f''$  of $\PO$ in rounds other than those rounds in $G'$ i which he grabs the pawns owning non-isolated vertices while playing against strategy $f'$ of \PO.
Then the resulting play in the always-grabbing game will still be losing for \PT, since otherwise, \PT could have followed the same order of grabbing the non-isolated pawns in the optional-grabbing game $\G'$ and would have won the optional-grabbing game.
This contradicts that $f'$ is winning for \PO in the optional-grabbing game.
This implies that if \PO has a winning strategy in the optional-grabbing game $\G'$ from the configuration $\zug{v,P}$, then he has a winning strategy in the always-grabbing game $\G''$ from the configuration $\zug{v, P \cup \{1, \dots, d+10\}}$.
% with the same set of initial pawns.

We now only need to show that in every round in $G''$ where \PT moves the token, \PT has an isolated pawn for \PO to grab.
We consider the following claim.
\begin{clm} \label{clm:tenPawns}
In the always-grabbing game $G''$, for strategy $f''$ of \PO and some strategy of \PT, every time the token is at a primed main vertex, that is, at the beginning of a $\delta$-path, \PT owns at least $10$ pawns owning the isolated vertices.
%If in the beginning of the game \PT owns $(k- |P_I|)+(k+10)$ pawns, then throughout the game, at every primed main vertex, we have that \PT owns at least $10$ isolated pawns.
\end{clm}
\begin{proof}
Consider the case in which at a primed main vertex $v$, \PT has $r$ pawns and the token is at $v$.
Now suppose that the token is moved along a $\delta$-path to the next primed main vertex $v'$. 
Note that at this next primed vertex $v'$, \PT has at least $r$ pawns. 
This clearly holds since by 
% Remark~\ref{rem:PLi_grab} and Remark~\ref{rem:PLni_grab}, 
Corollary~\ref{cor:grab}, along the $\delta$-path, $\PO$ grabs at most $10$ pawns and \PT grabs at least 10 pawns.
Thus, after traversing a $\delta$-path, the number of pawns that \PT controls does not decrease.
Now, in a play, every time the token is at a primed main vertex in $\G''$, it goes along a $\delta$-path to reach the next primed main vertex again.
Now the initial vertex is a primed main vertex and \PT has $(d- |P|)+(d+10)$ pawns at the beginning.
Here $P$ is the initial set of pawns owned by \PO in optional-grabbing game $\G'$.
Thus, every time the token reaches a primed main vertex afterwards, $\PT$ has at least $(d- |P|)+(d+10)$ pawns.
Now since there are exactly $d$ non-isolated pawns, and $(d- |P|)+(d+10)\geq (d+10)$, we have that out of the pawns that \PT has at a primed main vertex, at least $10$ are isolated pawns.
Hence proved.
%Now since every play that simulates a APSPACE Turing machine is a concatenation of $\delta$-paths.
%Hence in the always-grab game $G''$, every time the token is at a primed main vertex, $\PT$ has at least $(k- |P_I|)+(k+10)$ pawns, since that is the number of pawns that she owned initially.
\end{proof}
Now to prove Lemma~\ref{lem:always-grab_optional-grab}, since by Corollary~\ref{cor:grab}, we know that \PO has a strategy in the always-grabbing game such that from a primed main vertex to reach the next primed main vertex, he needs to grab in at most 10 rounds, and since by Claim~\ref{clm:tenPawns}, \PT has at least $10$ pawns owning isolated vertices, whenever the token is at primed main vertex, in every round between primed main vertices where \PO needs to grab, \PT has an isolated pawn that \PO can grab.
Hence by the argument above, if \PO has a winning strategy in the optional-grabbing game $\G'$, we have that \PO also has a winning strategy in the always-grabbing game $\G''$.

Note that the proof of this direction is also analogous for \PT.
We show that if \PT has a winning strategy in $\G'$, then she also has a winning strategy in $\G''$.
In particular, from Corollary~\ref{cor:grab}, along a $\delta$-path, \PO grabs in at least $10$ rounds and \PT grabs in at most $10$ rounds.
Also, similar to Claim~\ref{clm:tenPawns}, we can show that every time the token is at a primed main vertex, \PO owns at least $10$ pawns owning isolated vertices.
This is because initially, \PO controls $|P|+d+10$ vertices and this invariant is maintained throughout the play whenever the token reaches a primed main vertex.
Now $|P|+d+10 \geq d+10$, and hence, at the beginning of a $\delta$-path, \PO controls at least $10$ pawns owning isolated vertices.
% \end{proof}

% \begin{proof}[Proof of Lemma~\ref{lem:always-grab_optional-grab_iff}]
For the converse direction in Lemma~\ref{lem:always-grab_optional-grab}, we use the determinacy argument as follows.
Again, we show the proof for \PO that if \PO wins in the always-grabbing game $\G''$, then he also wins the optional-grabbing game $\G'$.
The proof for \PT is analogous.

Let \PO has a winning strategy in $\G''$.
Then, by determinacy, \PT loses in $\G''$.
Now from Lemma~\ref{lem:always-grab_optional-grab}, by taking the contrapositive for \PT, we have that \PT also loses $\G'$.
Hence, again by determinacy of pawn games, we have that \PO wins $\G'$.
\end{proof}

We now state the following theorem.
While the lower bound follows from Thm.~\ref{thm-MVPP optional grab} and Lem.~\ref{lem:always-grab_optional-grab}, the upper bound follows from Thm.~\ref{thm:exp}.
% The following theorem follows from Thm.~\ref{thm-MVPP optional grab} and Thm.~\ref{thm:exp}. 

% Thus we have the following theorem.
\begin{thm}
\label{thm-MVPP always grab}
MVPP always-grabbing PAWN-GAMES is $\EXP$-complete.
\end{thm}
% \begin{proof}
% The $\EXP$-hardness follows from Lemma~\ref{lem:always-grab_optional-grab_iff} since \PO has a winning strategy in the always-grab game $G''$ iff \PO has a winning strategy in the optional-grabbing game $G'$ which in turn simulates an alternating $\PSPACE$ Turing machine.

% The $\EXP$-membership follows since the configuration graph of the always-grab pawn game is exponential in size of the input game, and reachability game in a graph is in $\PTIME$.
% \end{proof}

We conclude this section by adapting Thm.~\ref{thm:optional-grabbing-detrimental} to always-grabbing. Namely, we show that adding pawns to a player is never beneficial in MVPP always-grabbing games (with the exception of the pawn that owns the current vertex).
% The proof can be found in App.~\ref{app:always-grabbing-detrimental}.

\begin{thm}
\label{thm:always-grabbing-detrimental}
Consider a configuration $\zug{v, P}$ of an MVPP always-grabbing pawn game $\G$. Let $j \in [d]$ such that $v \in V_j$ and $P' \subseteq P$. Assuming that $j \in P$ implies $j \in P'$, if \PO wins from $\zug{v, P}$, he wins from $\zug{v, P'}$. Assuming that $j \in \overline{P'}$ implies $j \in \overline{P}$, if \PT wins from $\zug{v, P'}$, she wins from $\zug{v, P}$. 
\end{thm}
\begin{proof}
We show the case when \PO has a winning strategy.
The case for \PT having a winning strategy is analogous.
Recall from the proof of Thm.~\ref{thm:optional-grabbing-detrimental} that the cases were argued for both configuration vertices and intermediate vertices.
For configuration vertices, The proof of this theorem is exactly the same as in the proof of Thm.~\ref{thm:optional-grabbing-detrimental}.
We detail below the case for intermediate vertices in the case of always-grabbing games.
We use the same notations as in the proof of Thm.~\ref{thm:optional-grabbing-detrimental}.

As in the proof of Thm.~\ref{thm:optional-grabbing-detrimental}, we again distinguish between two cases: $j \in P$ and $j \notin P$.
Consider an intermediate vertex $\zug{u, c} \in W_{n+1}$.
We will show that $\zug{u, c'} \in W_{n+1}$.
We denote by $\ell$, the pawn that owns $u$. 

First consider $j \in P$, thus \PO controls $c$.
We have as in the proof of Thm.~\ref{thm:optional-grabbing-detrimental} that both $\zug{u, c}$ and $\zug{u, c'}$ are \PT vertices.
Consider a neighbor of $\zug{u, c'}$, a configuration vertex $\zug{u,Q'}$. 
Note that in always-grabbing \PT has to grab a pawn $r$ from \PO, thus $Q'=P' \setminus \set{r}$.
There are two cases, either $\ell \in Q'$ or $\ell \notin Q'$.
Recall that in order to apply the induction hypothesis on $\zug{u, Q'}$, we need to find a neighbor $\zug{u, Q}$ of $\zug{u, c}$ such that if $\ell \notin Q'$, then $\ell \notin Q$.
If $\ell \notin Q'$, then we set $Q = P \setminus \{\ell\}$ if $\ell \in P$ and we set $Q = P \setminus \{r\}$ when $\ell \notin P$.
If $\ell \in Q'$, we set $Q = P \setminus \{r\}$.
% Note that in this case, we have that $r \neq \ell$.
The rest of the argument is as in the proof of Theorem~\ref{thm:optional-grabbing-detrimental}.

Next consider the case when $j \notin P$.
In this case, both $\zug{u, c}$ and $\zug{u, c'}$ are \PO vertices.
In always-grabbing, \PO grabs a pawn $r$ in $\zug{u, c}$, thus we have $Q = P \cup \{r\}$. % \supsetneq P$.
We find a neighbor $\zug{u, Q'}$ of $\zug{u, c'}$ to apply the induction hypothesis on.
If $\ell \in P$ and $\ell \in P'$, then we set $Q' = P' \cup \{r\}$.
If $\ell \in P$ and $\ell \notin P'$, then we set $Q' = P' \cup \{\ell\}$.
If $\ell \notin P$ and $\ell \notin P'$, then we set $Q' = P' \cup \{r\}$.
Again, the remaining argument is as in the proof of Theorem~\ref{thm:optional-grabbing-detrimental} and we are done.    
\end{proof}

% \newpage
% \section{grab or give game}
\section{Always Grabbing-or-Giving Pawn Games} \label{sec:always-grab-or-give}
In this section, we show that MVPP always grabbing or giving games are in PTIME. We find it intriguing that a seemingly small change in the mechanism -- allowing a choice between grabbing and giving instead of only grabbing -- reduces the complexity to $\PTIME$ from $\EXP$-complete.
We make the following simple observation.

\begin{obs} \label{obs:grab-or-give}
In an always grabbing or giving game, every time \PLi makes a move from a vertex $v$ to a vertex $u$, \PLni can decide which player controls $u$.
\end{obs}
If \PLni does not control $p_{u}$ that owns $u$ and he wants to control $u$, he can grab $p_{u}$ from \PLi.
If he does not want to control $u$ and if he has $p_{u}$, he can give it to \PLi.

Consider an always-grabbing-or-giving game $\G = \zug{V, E, T}$ and an initial configuration $c$. We construct a turn-based game $\G'$ and an initial vertex $v_0$ so that \PO wins in $\G$ from $c$ iff he wins in $\G'$ from $v_0$. Let $\G'=\zug{V', E', T'}$, where $V' = \{\zug{v,i}, \zug{\widehat{v},i} \:|\: v \in V, i \in \{1,2\}\}$ with $V_1'=\{\zug{v,1}, \zug{\widehat{v},1} \:|\: v \in V\}$ and $V_2'=\{\zug{v,2}, \zug{\widehat{v},2} \:|\: v \in V\}$, $T' = T \times \{1,2\}$, and $E' = \{(\zug{v,i}, \zug{\widehat{u},3-i}), (\zug{\widehat{u},3-i}, \zug{u,i}), (\zug{\widehat{u},3-i}, \zug{u,3-i}) \:|\: (v,u) \in E, i \in \{1,2\}\}$.
We call each vertex $\zug{v,i}$ a \emph{main} vertex and each $\zug{\widehat{v},i}$ an \emph{intermediate} vertex.
Suppose that \PLi moves the token from $v$ to $u$ in $\G$.
If \PLni decides to control $u$, then in $\G'$, the token moves from the main vertex $\zug{v,i}$ to the main vertex $\zug{u, 3-i}$, else from $\zug{v,i}$ to the main vertex $\zug{u, i}$, and in each case, through the intermediate vertex $(\widehat{u}, 3-i)$ that models the decision of \PLni on the control of $u$. The target vertices $T'$ are main vertices.
We can prove the following lemma.
% The proof of the following lemma appears in App.~\ref{app:grab-or-give}.
\begin{lem} \label{lem:always-grab-or-give}
%shibashis-cr : Uncommented the old version. The new version is incomplete since we need if and only if.
%\PLi has a winning strategy from configuration $\zug{v,P}$ in $\G$ with \PLi (\PLni) controlling $v$ iff \PLi has a winning strategy from $\zug{v,i}$ ($\zug{v,3-i}$) in $\G'$.
Suppose \PO wins from configuration $\zug{v,P}$ in $\G$. If he controls $v$, he wins from $\zug{v,1}$ in $\G'$, and if \PT controls $v$, \PO wins from $\zug{v,2}$ in $\G'$. Dually, suppose that \PT wins from $\zug{v, P}$ in $\G$. If she controls $v$, then she wins from $\zug{v, 2}$ in $\G'$, and if \PO controls $v$, \PT wins from $\zug{v, 1}$ in $\G'$.
\end{lem}
\begin{proof}
Suppose \PLi has a winning strategy in the game $\G$ from configuration $\zug{v,P}$.
We show that \PLi has a winning strategy in the game $\G'$ from vertex $\zug{v,i}$.
The other direction of the lemma follows from determinacy of two-player reachability games.
We prove the above for \PO.
The proof for \PT is analogous.

Let $W_j$ be the set of configurations in $\G$ such that \PO has a strategy to reach $T$ in at most $j$ rounds.
Let $A_j$ be the set of vertices in $\G'$ such that \PO has a strategy to reach $T'$ in at most $j$ rounds. We prove the following claim.
\begin{clm}
If $\zug{v,P} \in W_j$, with \PO(\PT) controlling $v$ then $\zug{v, 1}$ ($\zug{v, 2}$) belongs to $A_{2j}$.
\end{clm}
\begin{proof}
We prove this by induction on $j$. For the base case with $j=0$. this clearly holds.
Suppose the claim holds for $j=h$, and we now show that the claim holds for $j=h+1$.

Consider a configuration $\zug{v,P} \in W_{h+1}$.
We first look at the case when \PO controls $v$. 
We show that $\zug{v, 1} \in A_{2(h+1)}$.
Note that, by definition of $\G$ and $\G'$, both $\zug{v,P}$ and $\zug{v,1}$ are controlled by \PO in $\G$ and $\G'$ respectively.

Since $\zug{v,P} \in W_{h+1}$ and \PO controls vertex $v$, there is a strategy of \PO that takes the token starting from $v$ with pawns $P$ to a target vertex in $T$ in at most $h+1$ steps.
Let under this strategy, \PO moves the token from $v$ to $u$.
% Since $\zug{v,P} \in W_{h+1}$, there is a neighbour $u \in N(v)$ such
Note that $\zug{u, P'}\in W_h$ for all configurations $\zug{u, P'}$ that \PT can reach after grabbing from or giving \PO a pawn after \PO moves the token from $v$ to $u$.
Let $P_u$ be the pawn owning vertex $u$.
Now from Observation \ref{obs:grab-or-give}, we know that once \PO moves the token from vertex $v$
% configuration $\zug{v,P}$ 
to a vertex $u$, then \PT can reach a configuration $\zug{u, P'}$ with $p_u \in P'$ as well $\zug{u, P''}$ with $p_u \notin P''$. 
Thus there exist configurations $\zug{u,P'},\zug{u,P''}$ in $W_h$ such that $p_u\in P'$ and $p_u\notin P''$.
Now suppose the token is at vertex $\zug{v,1}$ in $\G'$.
Since $u$ is a neighbour of $v$, by the definition of $\G'$, we have that vertex $\zug{\widehat{u},2}$ is a neighbour of $\zug{u,1}$ in $\G'$. 
We show that $\zug{\widehat{u},2}\in A_{2h+1}$.
Recall that vertex $\zug{\widehat{u},2}$ is controlled by \PT and the only neighbours of this vertex are $\zug{u,1}$ and $\zug{u,2}$.
Now if we show that both $\zug{u,1}$ and $\zug{u,2}$ are in $A_{2h}$, then we are done.
%Now since we know that $\exists \zug{u,P'},\zug{u,P''}\in W_h$ such that $p_u\in P'$ and $p_u\notin P''$ by induction hypothesis $\zug{u,1},\zug{u,2}\in A_{2h}$.
Since we know that there exists $\zug{u,P'}$ in $W_h$ such that $p_u\in P'$, by the induction hypothesis, we have that $\zug{u,1}\in A_{2h}$, and similarly since there exists $\zug{u,P''}$ in $W_h$ such that $p_u\notin P''$, we have that $\zug{u,2}\in A_{2h}$.
%Since $u$ is a neighbour of $v$, note that in $\G'$ that the token first moves to $\zug{\widehat{u},2}$ from which \PT can move to either $\zug{u,1}$ or $\zug{u,2}$.
%Since $\zug{v,P} \in W_{h+1}$, there is a neighbour $u \in N(v)$ such that $\zug{u, P'}\in W_h$ for all configurations $\zug{u, P'}$ that \PT can reach after grabbing from or giving \PO a pawn.
%Let $p_u$ be the pawn owning the vertex $u$.
%Note that \PT can reach a configuration $\zug{u, P'}$ with $p_u \in P'$ as well $p_u \notin P'$.

The case for which \PO does not control $v$ can also be proved similarly.
\end{proof}
%shibashis-postcr
We show that \PLi wins in $\G$ from $\zug{v,P}$ when \PLi (\PLni) controls $v$ implies that \PLi wins from $\zug{v,i}$ ($\zug{v,3-i}$) in $\G'$.
By taking contrapositive, we have that \PLi loses from $\zug{v,i}$ ($\zug{v,3-i}$) in $\G'$ implies that \PLi loses in $\G$ from $\zug{v,P}$ when \PLi (\PLni) controls $v$.
By determinacy, we have that \PLni wins from $\zug{v,i}$ ($\zug{v,3-i}$) in $\G'$ implies that \PLni wins in $\G$ from $\zug{v,P}$ when \PLi (\PLni) controls $v$.

Hence showed.    
\end{proof}

Since the size of $\G'$ is polynomial in the size of $\G$, Thm.~\ref{thm:turn-based} implies the following. 

\begin{thm}
\label{thm:always-grab-or-give}
MVPP always-grab-or-give PAWN-GAMES is in PTIME.
\end{thm} 
% \begin{proof}
% We will reduce the Grab or Give pawn game to a simple game in no more than polytime.
% As a simple reachablity game can be solved in PTIME, Grab or Give pawn game with reachablity objective can be solved in PTIME.
% \end{proof}

%%%%%%%%%%%%%%%%%%%%%%%%%%%%%%%%%%%%%%%%%
\section{k-Grabbing Pawn Games}\label{sec:k-Grabbing Pawn Games}

In this section, we consider pawn games under $k$-grabbing in increasing level of generality of the mechanisms. We start with positive news.

%: a $\PTIME$ algorithm for reachability OVPP games. We show that safety OVPP games under a mild constraint on the mechanism are $\NP$-hard. We then study OMVPP games and show that they are $\PSPACE$-complete. 

\begin{thm}
\label{thm:OVPP-k-grabbing}
OVPP $k$-grabbing PAWN-GAMES is in $\PTIME$.
%Consider an OVPP $k$-grabbing game $\G$, with \PO grabbing, and an initial configuration $c$. There is a $\PTIME$ algorithm that decides whether \PO wins $\G$ from $c$.
\end{thm}
\begin{proof}
Let $k \in \Nat$, an OVPP $k$-grabbing game $\G = \zug{V, E, T}$, and an initial configuration $c = \zug{v_0, P_0}$, where we refer to $P_0$ as a set of vertices rather than pawns. Our algorithm solves a harder problem. It computes a {\em minimum-grabbing} function $\eta: V \rightarrow \Nat$ that labels each vertex $u \in V$ with the necessary and sufficient number grabs needed from $u$ to win. Formally, $\eta(u)$ is such that \PO wins an $\eta(u)$-grabbing game played on $\G$ from configuration $\zug{u, P_0}$ but loses an $\big(\eta(u) - 1\big)$-grabbing game from configuration $\zug{u, P_0}$. The algorithm is depicted in Alg.~\ref{alg:k-grabbing}. It calls the $\Call{Solve-Turn-Based-Game}$, which returns the set of vertices that are winning for \PO in a turn-based reachability game, e.g., using attractor-based computation as in Thm.~\ref{thm:turn-based}. 

\begin{algorithm}[t]
\begin{algorithmic}[1]
\State $\ell = 0$
\While {$\exists u \in V$ that is not labeled by $\eta$}
\State $W^1_\ell = \Call{Solve-Turn-Based-Game}{V_1= P_0, V_2 = V \setminus P_0, E, T}$
\State Define $\eta(u) = \ell$, for all $u \in W^1_\ell$ that is not yet labeled. 
\State $B_\ell = \set{u \in V \setminus W^1_\ell: N(u) \cap W^1_\ell \neq \emptyset}$
\State $T = B^\ell$ and $\ell = \ell+1$.
\EndWhile
\end{algorithmic}
    \caption{\label{alg:k-grabbing}
    Given an OVPP pawn game $\G = \zug{V, E, T}$ and a set of pawns $P_0 \subseteq V$ that \PO controls, the algorithm returns a minimum-grabbing function $\eta: V \rightarrow \Nat$.}
\end{algorithm}

For the base case, consider the turn-based game $\G_0 = \zug{V, E, T}$ with $V_1 = P_0$. Let $W^1_0 \subseteq V$ denote \PO's winning region in $\G_0$. Clearly, for every $u \in W^1_0$, we have $\eta(v_0) =0$, and for every $u \notin W^1_0$, we have $\eta(v_0) \geq 1$. For the inductive step, suppose that for $\ell \geq 0$, the set $W^1_\ell = \set{u \in V: \eta(u) \leq \ell}$ has been found. That is, for every $u \notin W^1_\ell$, \PT has a strategy that wins the $\ell$-grabbing pawn game $\G$ from configuration $\zug{u, P_0}$. We show how to find $W^1_{\ell+1}$ in linear time. Let the {\em border} of $W^1_\ell$, denoted $B_\ell$, be the set of vertices from which $W^1_\ell$ can be reached in one step, thus $B_\ell = \set{v \in V:v\notin W^1_\ell : N(v) \cap W^1_\ell \neq \emptyset}$.
Note that the vertices in $B_\ell$ are all controlled by \PT since otherwise, such vertices will be in the set $W^1_\ell$.
% In App.~\ref{app:OVPP-k-grabbing}, 
Now we show that a vertex $u \notin W^1_\ell$ has $\eta(u) = \ell+1$ iff \PO can force the game from configuration $\zug{u, P_0}$ to a vertex in $B_\ell$ in one or more rounds without making any grab. 
% and in 
% a {\em non-trivial} manner. 
\PO wins from such a vertex $u$ by forcing the game into $B_\ell$, grabbing the pawn in $B_\ell$, and proceeding to $W_\ell$, where by the induction hypothesis, he wins with the remaining grabs. 
Computing $W^1_{\ell+1}$ roughly entails a solution to a turn-based game with target set $B_\ell \cup W^1_\ell$.

Intuitively, we show that a vertex $u \notin W^1_\ell$ has $\eta(u) = \ell+1$ iff \PO can force the game from configuration $\zug{u, P_0}$ to a vertex in $B_\ell$ without making any grabs and in a {\em non-trivial} manner. We compute $W^1_{\ell+1}$ as follows. Consider the turn-based game $\G_\ell = \zug{V, E, B_\ell \cup W^1_\ell}$ with $V_1 = P_0$. We say that \PO wins non-trivially from $u \in V$ if he has a strategy $f$ that guarantees the following against any \PT strategy: the resulting play is of the form $u=v_0, v_1, v_2,\ldots, v_m$ with $v_m \in (B_\ell \cup W^1_\ell)$ and it is non-trivial, meaning that $m \geq 1$. Note the difference from standard turn-based games: a target vertex $u \in B_\ell \cup W^1_\ell$ might be losing for \PO when considering winning non trivially, e.g., when $u$ has an outgoing edge to a sink, no non-trivial play that starts in $u$ ends in a target vertex. 

Let $U^1_\ell$ denote the set of vertices from which \PO wins in $\G_\ell$ non-trivially. We describe an algorithm for finding $U^1_\ell$. 
We construct a game $\G'_\ell$ in which we make a copy $u'$ of each vertex $u \in B_\ell$, and set the neighbors of $u'$ to be $\{v\in N(u): v\notin W^1_\ell \}$. Intuitively, if \PO wins from $u'$ in $\G'_\ell$, he can win in $\G_\ell$ from $u$ in a non-trivial manner: since $u'$ is not winning in $\G'_\ell$, \PO must make at least one step before entering $B_\ell$.
Let $U'$ be the winning set for \PO in $\G'_\ell$. 
It is not hard to see that $U^1_\ell = \set{u \notin B_\ell: u \in U'} \cup \set{u \in B_\ell: u' \in U'}$. 

Let $u \in U^1_\ell$. We claim that if $u\notin W_{\ell}^1$ then $\eta(u) = \ell+1$. First, since $u\notin W_{\ell}^1$, by the induction hypothesis, $\eta(u) \geq \ell+1$. Next, we show that $\eta(u) = \ell+1$ by showing that \PO wins the $(\ell+1)$-grabbing pawn game $\G$ from configuration $\zug{u, P_0}$. \PO initially plays according to a strategy that forces the game to $B_\ell$ non-trivially without grabbing. 
Suppose that \PT is playing according to some strategy and the game visits $u'$ before visiting $u'' \in B_\ell$. Then, \PO grabs $u''$, proceeds to $W^1_\ell$, and uses his winning $\ell$-grabbing strategy from there. 

Let $u \notin U^1_\ell \cup W^1_\ell$. We claim that $\eta(u) > \ell+1$. Note that in order to reach $W^1_\ell$ and in particular reach $T$, the game must first visit $B_\ell$. Suppose that \PT is following a winning strategy in $\G_\ell$.  Thus, in order to reach $B_\ell$, \PO must grab $u' \notin B_\ell \cup W^1_\ell$. Suppose that the game reaches a configuration $\zug{u'', P_0 \cup \set{u'}}$, where $u'' \in B_\ell$. By the above, $\eta(u'') \geq \ell+1$. That is, any \PO strategy that wins from $u''$ must make at least $\ell+1$ grabs. Hence, in order to win from $u$, \PO makes at least $\ell+2$ grabs. 

Let $n = |V|$. Note that $W^1_n = V$. Computing $W^1_{\ell+1}$ from $W^1_\ell$ requires a call to an algorithm that solves a reachability game, which can be done in linear time. Hence, the total running time of the algorithm is polynomial in the size of $\G$.
\end{proof}

\makeatletter
\def\squarecorner#1{
    \pgf@x=\the\wd\pgfnodeparttextbox%
    \pgfmathsetlength\pgf@xc{\pgfkeysvalueof{/pgf/inner xsep}}%
    \advance\pgf@x by 2\pgf@xc%
    \pgfmathsetlength\pgf@xb{\pgfkeysvalueof{/pgf/minimum width}}%
    \ifdim\pgf@x<\pgf@xb%
        \pgf@x=\pgf@xb%
    \fi%
    \pgf@y=\ht\pgfnodeparttextbox%
    \advance\pgf@y by\dp\pgfnodeparttextbox%
    \pgfmathsetlength\pgf@yc{\pgfkeysvalueof{/pgf/inner ysep}}%
    \advance\pgf@y by 2\pgf@yc%
    \pgfmathsetlength\pgf@yb{\pgfkeysvalueof{/pgf/minimum height}}%
    \ifdim\pgf@y<\pgf@yb%
        \pgf@y=\pgf@yb%
    \fi%
    \ifdim\pgf@x<\pgf@y%
        \pgf@x=\pgf@y%
    \else
        \pgf@y=\pgf@x%
    \fi
    \pgf@x=#1.5\pgf@x%
    \advance\pgf@x by.5\wd\pgfnodeparttextbox%
    \pgfmathsetlength\pgf@xa{\pgfkeysvalueof{/pgf/outer xsep}}%
    \advance\pgf@x by#1\pgf@xa%
    \pgf@y=#1.5\pgf@y%
    \advance\pgf@y by-.5\dp\pgfnodeparttextbox%
    \advance\pgf@y by.5\ht\pgfnodeparttextbox%
    \pgfmathsetlength\pgf@ya{\pgfkeysvalueof{/pgf/outer ysep}}%
    \advance\pgf@y by#1\pgf@ya%
}
\makeatother
\pgfdeclareshape{square}{
    \savedanchor\northeast{\squarecorner{}}
    \savedanchor\southwest{\squarecorner{-}}

    \foreach \x in {east,west} \foreach \y in {north,mid,base,south} {
        \inheritanchor[from=rectangle]{\y\space\x}
    }
    \foreach \x in {east,west,north,mid,base,south,center,text} {
        \inheritanchor[from=rectangle]{\x}
    }
    \inheritanchorborder[from=rectangle]
    \inheritbackgroundpath[from=rectangle]
}

\begin{figure}[t]
\centering
% \scalebox{0.7}{
\begin{tikzpicture}[node distance={8mm},initial text = ,main/.style = {draw,circle, text=black, inner sep=1pt},main1/.style = {draw,circle, blue, left color=blue!20, text=black, inner sep=1pt},main2/.style = {draw,circle, red, left color=red!20, text=black, inner sep=1pt},main3/.style = {draw,circle, green, left color=green!20, text=black, inner sep=1pt},gain1/.style = {draw,square, blue, left color=blue!20, text=black, inner sep=1pt},gain2/.style = {draw,square, red, left color=red!20, text=black, inner sep=1pt},gain3/.style = {draw,square, green, left color=green!20, text=black, inner sep=1pt}] 
\node[main,initial] (1) {1};
\node[gain1] (2) [above right of=1] {$S_1^1$};
\node[gain2] (3) [below right of=1] {$S_2^1$};
\node[main] (4) [below right of=2] {2};
\node[gain3] (5) [below right of=4] {$S_3^2$};
\node[gain2] (6) [above right of=4] {$S_2^2$};
\node[main] (7) [above right of=5] {$3$};
\node[gain3] (8) [right of=7] {$S_3^3$};
\node[main,accepting] (9) [right of=8] {$t$};
\node[main] (10) [right of=6] {$s$};
\node[main] (11) [right of=5] {$s$};
\draw[->] (1) -- (2);
\draw[->] (1) -- (3);
\draw[->] (2) -- (4);
\draw[->] (3) -- (4);
\draw[->] (4) -- (5);
\draw[->] (4) -- (6);
\draw[->] (6) -- (7);
\draw[->] (5) -- (7);
\draw[->] (7) -- (8);
\draw[->] (2) to [out=90,in=90,looseness=0.5] (10);
\draw[->] (6) -- (10);
\draw[->] (3) to [out=270,in=270,looseness=0.5] (11);
\draw[->] (5) -- (11);
\draw[->] (8) -- (10);
\draw[->] (8) -- (9);
\end{tikzpicture}
% }
%\end{minipage}
\caption{Consider the input to SET-COVER $U = [3]$ and $\S = \set{\set{1}, \set{1,2}, \set{2,3}}$. 
The figure depicts the output on this input of the reduction in 
% reductions in Thm.~\ref{thm:safety-k-grabbing} (left) and 
Thm.~\ref{thm:reach-MVPP-k-grabbing}}
% (right).} 
\label{fig:NP-hard}
\end{figure}
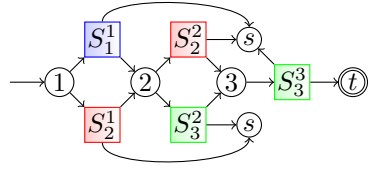
The proof of the following theorem, which 
% can be found in App.~\ref{app:reach-MVPP-k-grabbing}, 
is obtained by a reduction from SET-COVER. 
\begin{thm}
\label{thm:reach-MVPP-k-grabbing}
MVPP $k$-grabbing game PAWN-GAMES is NP-hard.
\end{thm}
\begin{proof}
Given an input $\zug{U, \S, k}$ to SET-COVER, we construct an MVPP $k$-grabbing pawn game $\G = \zug{V, E, T}$ in which \PO grabs (see Fig.~\ref{fig:NP-hard}). Intuitively, $\G$  has a ``chain-like'' structure, and in order to win, \PO must cross the chain. Certain positions in $\G$ corresponds to $i \in U$. Each neighbor of $i$ correspond to a set $S \in \S$ with $i \in S$. Thus, a choice of \PO at $i$ can be thought of as assigning a set in $\S$ that covers $i$. We construct $\G$ so that before moving to a neighbor $S$ of $i$, \PO must grab the pawn that owns $S$.  
All vertices that correspond to $S$ are controlled by the same pawn, thus if \PO moves from $i' > i$ to a neighbor that corresponds to $S$, there is no need to grab again. Since \PO is allowed only $k$ grabs, there is a one-to-one correspondence between winning \PO strategies and set covers of size $k$.

We describe the construction of $\G$ formally. Let $V = U \cup (\S \times U) \cup \set{s, t}$. \PO's target is $t$ and $s$ is a vertex with no path to $t$, thus it can be thought of as a target for \PT. There are $m+1$ pawns. For $j \in [m]$, Pawn~$j$ owns all vertices in $\set{\zug{S_j, i}: i \in U}$. Pawn~$0$ owns the vertices in $U$. We describe the edges. For $i \in U$, we define $N(i) = \set{\zug{S_j, i}: i \in S_j}$. For $1 \leq i \leq n-1$ and $j \in [m]$, we define $N(\zug{S_j, i}) = \set{i+1, s}$ and $N(\zug{S_j, n}) = \set{t, s}$. That is, if \PO moves from $i$ to its neighbor $S_j$ without controlling Pawn~$j$, then \PT will proceed to $s$ and win the game. 

We claim that a set cover $\S'$ of size $k$ gives rise to a winning $k$-grabbing \PO strategy. Indeed, by choosing, for each $i \in U$, to move to $S \in \S'$ and grabbing it if it has not been grabbed previously, \PO guarantees that $t$ is reached using at most $k$ grabs. On the other hand, observe a play of a winning $k$-grabbing \PO strategy against a {\em reasonable} \PT strategy; namely, a strategy that always moves to $s$ from a neighboring vertex $u$ when \PT controls $u$. Suppose that the set of pawns $\S'$ is grabbed. It is not hard to see that $\S'$ is a set cover of size at most $k$, and we are done.
\end{proof}

%short \subsection{OMVPP $k$-grabbing pawn games}
We conclude this section by studying OMVPP games.% and show that they are PSPACE-complete.

\begin{lem}
\label{lem:OMVPP-PSPACE-h}
OMVPP $k$-grabbing PAWN-GAMES is PSPACE-hard.
\end{lem}
\begin{proof}
Consider an input $\phi = Q_1 x_1 \ldots Q_n x_n C_1 \wedge \ldots \wedge C_m$ to TQBF, where $Q_i \in \set{\exists, \forall}$, for $1 \leq i \leq n$, each $C_j$, for $1 \leq j \leq m$, is a clause over the variables $x_1, \ldots, x_n$. We construct an OMVPP $n$-grabbing pawn game $\G = \zug{V, E, T}$ such that \PO wins iff $\phi$ is true. 

The intuition is similar to Thm.~\ref{thm:reach-MVPP-k-grabbing}. We construct $\G$ to have a chain-like structure that \PO must cross in order to win. The chain consists of two parts. In the first part, each position corresponds to a variable $x_i$, for $i \in [n]$. We construct $\G$ so that if $x_i$ is existentially quantified, then \PO controls $x_i$, and if $x_i$ is universally quantified, then \PT controls $x_i$ and \PO cannot win by grabbing it. We associate a move in $x_i$ with an assignment to $x_i$. Technically, $x_i$ has two neighbors $v_i$ and $\neg v_i$, initially at the control of \PT. If the token reaches a neighbor of $x_i$ without \PO grabbing it, then \PO loses the game. \PO is allowed $n$ grabs, thus in order to win, it is necessary for him to grab, for each $i$, one of the neighbors of $x_i$. It follows that a play that crosses the first part of the chain gives rise to an assignment $f$ to $x_1,\ldots, x_n$. 

In the second part of $\G$, we verify that $f$ is valid. Each position of $\G$ corresponds to a clause $C_j$, for $j \in [m]$, all of which are at the control of \PT in the initial configuration of $\G$. Note that once the first part of $\G$ is crossed,  \PO is not allowed any more grabs. The idea is that when arriving at $C_j$, for $j \in [m]$, \PO must control $C_j$, otherwise he loses. Recall that in OMVPP, it suffices to control one of the pawns that owns $C_j$ in order to control it. We define the pawns that own $C_j$ as follows. For $i \in [n]$, call $p_i$ the pawn that owns $v_i$ and $\neg p_i$ the pawn that owns $\neg v_i$. Thus, grabbing $p_i$ and $\neg p_i$ respectively corresponds to an assignment that sets $x_i$ to true and false. If $x_i$ appears in $C_j$, then $p_i$ is an owner of $C_j$, and if $\neg x_i$ appears in $C_j$, then $\neg p_i$ is an owner of $C_j$. Thus, \PO controls $C_j$ iff $f(C_j) = \text{true}$. It follows that \PO wins iff $f$ is valid. 

Formally, the vertices of $\G$ are $V = \set{s, t} \cup \set{x_i, v_i, \neg v_i: 1 \leq i \leq n} \cup \set{C_j: 1 \leq j \leq m}$.
The target vertex is $t$. The vertex $s$ is a sink vertex that is winning for \PT, i.e., there is no path from $s$ to $t$. The game consists of $2n+2$ pawns. For each $1 \leq i \leq n$, we associate with variable $x_i$ two pawns, which we denote by $p_i$ and $\neg p_i$. Intuitively, we will construct the game such that in order to win, \PO must grab either $p_i$ or $\neg p_i$, and we associate grabbing $p_i$ and $\neg p_i$ with an assignment that sets $x_i$ to true and false, respectively. Formally, if $x_i$ appears in $C_j$, then $p_i$ is an owner of $C_j$, and if $\neg x_i$ appears in $C_j$, then $\neg p_i$ is an owner of $C_j$. 
We use~$1$ and~$2$ to refer to the final two pawns. 
Pawn~$1$ owns every vertex $x_i$ such that variable $x_i$ is existentially quantified in $\phi$ and Pawn~$2$ owns every vertex $x_i$ such that $x_i$ is universally quantified in $\phi$. In the initial configuration, \PO controls only Pawn~$1$.

We describe the edges in the game, which has a chain-like structure. The last vertex on the chain is the target $t$, thus \PO must cross the chain in order to win.
We assume that \PT follows a {\em reasonable} strategy; namely, a strategy that proceeds to win at $s$ given the option, i.e., when the game reaches a vertex $u$ at her control and that neighbors $s$. 
The token is initially placed on $x_0$. For $1 \leq i \leq n$, the neighbors of $x_i$ are $v_i$ and $\neg v_i$. We associate with the choice at $x_i$, an assignment to $x_i$ in the expected manner. 
Suppose that the token is placed on $u \in \set{v_i,\neg v_i}$. We require \PO to grab the pawn that owns $u$ by adding an edge from $u$ to $s$. That is, since initially \PT controls the pawn that owns $u$, if \PO does not grab it, \PT will win from  $u$. For $1 \leq i < n$, the neighbor of $v_i$ and $\neg v_i$ is $v_{i+1}$. The neighbor of $v_n$ and $\neg v_n$ is $C_1$. 

Suppose that \PO follows a strategy that leads the token to $C_1$. We observe that since Pawn~$1$ is controlled by \PO, \PO chooses the assignment of the existentially-quantified variables. 
We claim that \PO's strategy does not grab Pawn~$2$, thus \PT chooses the assignment of the universally-quantified variables. 
Indeed, since \PO is restricted to grab at most $n$ pawns and must grab either $p_i$ or $\neg p_i$, for each $1 \leq i \leq n$. If he grabs Pawn~$2$, then there exists $1 \leq i \leq n$ such that he cannot grab any of $p_i$ or $\neg p_i$, and the game will necessarily reach either $v_i$ or $\neg v_i$ at the control of \PT, from which she wins.

We describe the outgoing edges from clause vertices. 
For $1 \leq j <m$, the vertex $C_j$ has two neighbors, $s$ and $C_{j+1}$, and the neighbors of $C_m$ are $s$ and $t$. That is, in order to draw the game to $t$, \PO must cross all clause vertices. Recall that the definition of OMVPP dictates that \PO chooses how to move the token at a vertex $u$ if he controls at least one of the pawns that owns $u$. Thus, a winning \PO strategy must grab, for each $1 \leq j \leq m$, at least one pawn that owns $C_j$. In turn, grabbing a pawn that owns $C_j$ means that the assignment to $x_1,\ldots,x_n$ that arises from the players' strategies satisfies $C_j$. It follows that \PO wins the game iff $\phi$ is true.
\end{proof}

We turn to study the upper bound. The following lemma bounds the provides a polynomial bound on the length of a winning play for \PO. The core of the proof intuitively shows that we can restrict attention to \PO strategies that grab at least once in a sequence of $|V|$ rounds. Otherwise, the game enters a cycle that is winning for \PT. 
We thus obtain a polynomial bound on the length of a winning play for \PO.

\begin{lem} \label{lem:stepsOMVPP}
Consider an OMVPP $k$-grabbing PAWN-GAME $\G = \zug{V, E, T}$, and an initial configuration $c$ that is winning for \PO. 
Then, \PO has a strategy such that, for every \PT strategy, a target in $T$ is reached within $|V| \cdot (k+1)$ rounds.
\end{lem}
\begin{proof}
Consider the turn-based game $\G'$ that corresponds to $\G$. Recall that $c$ is a vertex in $\G'$. Fix a \PO memoryless winning \PO strategy in $\G'$, which exists since $\G'$ is a turn-based reachability game. Consider some \PT strategy and let $\pi$ be the resulting play. We make three observations. (1) Each vertex in $\G'$ is visited at most once by $\pi$. Otherwise, $\pi$ enters a loop, which is losing for \PO. (2) The configurations can be partially ordered: a configuration $(v, P)$ can only be visited before a configuration $(u, P')$, for $P \subseteq P'$. Indeed, the only manner in which control of pawns changes is by \PO grabbing, which only adds pawns to the set that he already controls. (3) If the game starts from $(v, P)$, then it ends in a configuration $(u, P')$ with $|P'| \leq |P| + k$. Indeed, \PO can only grab $k$ times. Combining (1)-(3) implies that $\pi$ visits $T$ within $|V| \cdot (k+1)$ rounds.    
\end{proof}

For the upper bound, 
% in Appendix~\ref{app:k-grabbing-upper}, 
we describe an algorithm performing a depth-first traversal of the configuration graph of a game while storing, at a time, only a branch in $\PSPACE$. 
\begin{lem}
OMVPP $k$-grabbing PAWN-GAMES is in $\PSPACE$.
\end{lem}
\begin{proof}
We describe a $\PSPACE$ algorithm for OMVPP $k$-grabbing PAWN-GAMES as follows.
% The algorithm simulates a depth-first search (DFS) traversal of the configuration graph of the game.
% We will show that by Lemma~\ref{lem:stepsOMVPP} the depth of the tree is bounded by $n \cdot (k+1)$.
The algorithm explores an unwinding of the configuration graph of the $k$-grabbing pawn game in a depth-first manner.
We call this unwinding of the configuration graph a \emph{game tree}.
Recall that after each move by a player in the game, \PO chooses to grab a pawn and if \PO is controlling the current vertex, that is, where the token lies, then \PO chooses a successor, otherwise \PT chooses a successor.
A vertex $v$ that is controlled by \PO is an OR-vertex in the game tree in the sense that there should exist a successor of $v$ which accepts.
On the other hand, a vertex $v$ that is controlled by \PT is an AND-vertex in the game tree in the sense that all the successors of $v$ should accept.

Since by Lemma~\ref{lem:stepsOMVPP}, if \PO has a winning strategy then he has one such that for all strategies of \PT, he wins in $n \cdot (k+1)$ steps, 
% the depth of the DFS tree is at most $n \cdot (k+1)$.
it is sufficient to unwind the configuration graph such that the length of a path in the game tree starting from the initial vertex does not exceed $n \cdot (k+1)$.
At any time during exploring the game tree in a depth-first manner, the algorithm stores the path that is being currently traversed from the initial vertex, and the length of the path is thus at most $n \cdot (k+1)$.
In a depth-first traversal, from each vertex, its successors are visited in a particular order.
At each level of the path that has been currently traversed, the algorithm also keeps count of how many successors of the vertex at that level have been visited, and also the depth of the level from the root of the tree.
The latter ensures that the algorithm does not unwind the configuration graph to an extent so that the length of a path in the game tree exceeds $n \cdot (k+1)$.
Since in the configuration graph, there are at most exponentially many vertices of the form $\zug{v, P}$ where $v$ is a vertex of the $k$-grabbing pawn game and $P$ is a set of pawns, at each level, the count of number of successors that have been visited so far can be stored in $\PSPACE$.
Since the number of levels of the current path is bounded by $n \cdot (k+1)$ which is polynomial and each level uses polynomial space, the algorithm is in $\PSPACE$.
\end{proof}
By Lem.~\ref{lem:stepsOMVPP}, each branch of such a traversal has polynomial length, leading to the PSPACE upper bound.
We thus have the following.

\begin{thm}
\label{thm:PSPACE}
OMVPP $k$-grabbing PAWN-GAMES is PSPACE-complete.
\end{thm}

\begin{rem} \label{rem:kgrab_more_pawns_better}
We also note that in the case of $k$-grabbing mechanism, unlike optional-grabbing or always-grabbing mechanisms, it is always the case that for \PO, at every vertex, controlling more pawns is at least as good as controlling fewer pawns.
This is because, if \PO needs to control a particular vertex $v$ in any round of the game in order to win, he can grab the pawn controlling the vertex $v$ if the pawn does not belong to him regardless of which player moves the token provided he has not already grabbed $k$ pawns.
\end{rem}

\section{Discussion}
We introduce pawn games, a class of two-player turn-based games in which control of vertices  changes  dynamically throughout the game. Pawn games constitute a class of succinctly-represented turn-based games. %Still, 
We identify natural classes that are in PTIME. Our EXPTIME-hardness results are based on \LAK games, which we hope will serve as a framework for proving lower bounds. 
%short For example, a simple alternative proof that sabotage games are PSPACE-hard~\cite{LR03} can be shown based on a variant of \LAK games in which locks only close. 
We mention directions for future research. First, we leave several open problems; e.g., for MVPP $k$-grabbing pawn games, we only show NP-hardness and membership in PSPACE. Second, we focused on reachability games. It is interesting to study pawn games with richer objectives such as parity or quantitative objectives.
Quantitative objectives are especially appealing since one can quantify the benefit of controlling a pawn. 
Third, it is interesting to consider other pawn-transferring mechanisms and to identify properties of mechanisms that admit low-complexity results. Finally, grabbing pawns is a general concept and can be applied to more involved games like stochastic or concurrent games.

%%%%%%%%%%%%%%%%%%%%%%%%%%%%%%%%%%%%%%%%%%%%%%%%%%%%%%%%%%%%%%%%%%%%%%%%%%%%%%%%%%%%%%%%%%%
\bibliographystyle{alphaurl}
\bibliography{ga.bib}

\end{document}